\documentclass{ws-ijmpa}                                                                                                         
\usepackage[super,compress]{cite}
\usepackage{hyperref}
\hypersetup{colorlinks,urlcolor=black,citecolor=black,linkcolor=black,filecolor=black}
\usepackage{breakurl}
\usepackage{graphicx}
\usepackage{subfigure}
\usepackage{color}
\usepackage{float}
\usepackage{mathrsfs}

\newcommand{\bfg }{\begin{figure}[htpb]}
\newcommand{\efg }{\end{figure}}
\newcommand{\bmn }{\begin{minipage}}
\newcommand{\emn }{\end{minipage}}
\newcommand{\bt }{\begin{table}[htpb]}
\newcommand{\et }{\end{table}}

\newcommand{\pA}    {p+A}
\newcommand{\dA}    {d+A}
\newcommand{\pdA}   {p(d)+A}

\newcommand{\pdAu}  {p(d)+Au}
\newcommand{\AuAu}  {Au+Au}
\newcommand{\pPb}   {p+Pb}
\newcommand{\PbPb}  {Pb+Pb}
\newcommand{\Ru}	{^{96}_{44}\rm{Ru}}
\newcommand{\Zr}	{^{96}_{40}\rm{Zr}}
\newcommand{\RuRu}	{^{96}_{44}\rm{Ru} + ^{96}_{44}\rm{Ru}}
\newcommand{\ZrZr}	{^{96}_{40}\rm{Zr} + ^{96}_{40}\rm{Zr}}

\newcommand{\sNN}{$\sqrt{s_{\rm NN}}$ }

\newcommand{\GeVc}{GeV/$c$ }
\newcommand{\GeVcsq}{GeV/$c^2$ }

\newcommand{ \be }{\begin{equation}}
\newcommand{ \ee }{\end{equation}}
\newcommand{ \bea }{\begin{eqnarray}}
\newcommand{ \eea }{\end{eqnarray}}

\newcommand{ \pT} {p_{\rm T}}

\newcommand {\snn}  {\sqrt{s_{_{\rm NN}}}}

\newcommand {\pt}   {p_{T}}

\newcommand {\vexe} {v_{2,{\rm ebye}}}
\newcommand {\vobs} {\vexe^{\rm obs}}

\newcommand {\psiPP}    {\psi_{\rm PP}}
\newcommand {\psiRP}    {\psi_{\rm RP}}
\newcommand {\psiEP}    {\psi_{\rm EP}}
\newcommand {\phires}   {\phi_{\rm res}}

\newcommand {\vres} {v_{2,{\rm res}}}
\newcommand {\Bvec} {\vec{B}}

\newcommand {\gSS}  {\gamma_{\rm SS}}
\newcommand {\gOS}  {\gamma_{\rm OS}}
\newcommand {\gdel} {\Delta\gamma}
\newcommand {\dg}	{\Delta\gamma}
\newcommand {\dgscale}	{\dg_{\rm scaled}}

\newcommand {\mult}	{N}

\newcommand {\minv}	{m_{inv}}

\newcommand {\Cpm}[1]  {C_{\psi_{#1}}(\Delta{S})}
\newcommand {\Cpmp}[1] {C_{\psi_{#1}}^{\perp}(\Delta{S})}
\newcommand {\Rsm}[1]  {R_{\psi_{#1}}(\Delta{S})}

\newcommand {\mean}[1]  {\langle #1\rangle}
\newcommand {\note}[1]  {}

\begin{document}
\markboth{Jie Zhao}{Search for the Chiral Magnetic Effect in Relativistic Heavy-Ion Collisions}

\catchline{}{}{}{}{}
%
%%%%%%%%%%%%%%%%%%%%%%%%%%%%%%%%%%%%%%%%%%%%%%%%%%%%%%%%%%%%%%%%%%%%

\title{Search for the Chiral Magnetic Effect in Relativistic Heavy-Ion Collisions}

\author{Jie Zhao}

\address{Department of Physics and Astronomy, Purdue University,\\
West Lafayette, IN 47907, US\\
zhao656@purdue.edu}

%\author{Second Author}
%
%\address{Group, Laboratory, Address\\
%City, State ZIP/Zone, Country\\
%second\_author@domain\_name}

\maketitle

\begin{history}
\received{Day Month Year}
\revised{Day Month Year}
\end{history}

\begin{abstract}
Relativistic heavy-ion collisions provide an ideal environment to study the emergent phenomena in quantum chromodynamics (QCD). 
The chiral magnetic effect (CME) is one of the most interesting, arising from the topological charge fluctuations of QCD vacua, immersed in a strong magnetic field.
Since the first measurement nearly a decade ago of the possibly CME-induced charge correlation, extensive studies have been devoted to background contributions to those measurements.
Many new ideas and techniques have been developed to reduce or eliminate the backgrounds.
This article reviews these developments and the overall progress in the search for the CME. 

\keywords{Chiral magnetic effect; topological charge; heavy-ion collisions; QGP; QCD}
\end{abstract}

\ccode{PACS numbers: 25.75.-q, 25.75.Gz, 25.75.Ld, 25.75.Ag}

%25.75.-q   Relativistic heavy-ion collisions (collisions induced by light ions studied to calibrate relativistic heavy-ion collisions should be classified under both 25.75.-q and sections 13 or 25 appropriate to the light ions)
%25.75.Ag   Global features in relativistic heavy ion collisions
%25.75.Bh   Hard scattering in relativistic heavy ion collisions
%25.75.Cj   Photon, lepton, and heavy quark production in relativistic heavy ion collisions
%25.75.Dw   Particle and resonance production
%25.75.Gz   Particle correlations and fluctuations
%25.75.Ld   Collective flow
%25.75.Nq   Quark deconfinement, quark-gluon plasma production, and phase transitions (see also 12.38.Mh Quark-gluon plasma in quantum chromodynamics; 21.65.Qr Quark matter in nuclear matter)
%\PACS 25.75.-q \sep 25.75.Gz \sep 25.75.Ld \sep 25.75.Ag


%\tableofcontents

%%%%%%%%%%%%%%%%%%%%%%%%%%%%%%%%%%%%%%%%%%%%%%%%%%%%%%%%%%%%%%%%%%%%%%%%%%%%%%%%%%%%%%%%%%%%%%
\section{Introduction}
Quark interactions with topological gluon configurations can induce chirality imbalance and local parity violation 
in quantum chromodynamics (QCD)\cite{Lee:1973iz,Lee:1974ma,Morley:1983wr,Kharzeev:1998kz,Kharzeev:2004ey,Kharzeev:2007jp,Fukushima:2008xe,Kharzeev:2015znc}.
In relativistic heavy-ion collisions, this can lead to observable electric charge separation along the direction of the strong magnetic
field produced by spectator protons\cite{Kharzeev:2004ey,Kharzeev:2007jp,Fukushima:2008xe,Muller:2010jd,Kharzeev:2015znc}. 
This phenomenon is called the chiral magnetic effect (CME).
An observation of the CME-induced charge separation would confirm several fundamental properties of QCD, namely, approximate chiral symmetry restoration, 
topological charge fluctuations, and local parity violation. 
Extensive theoretical efforts have been devoted to characterize the CME, 
and intensive experimental efforts have been invested to search for the CME in heavy-ion collisions at BNL's Relativistic Heavy Ion Collider (RHIC) and CERN's Large Hadron Collider (LHC)\cite{Kharzeev:2015znc}.

Transitions between gluonic configurations of QCD vacua can be described by instantons/sphelarons and characterized by the Chern-Simons topological charge number.
Quark interactions with such topological gluonic configurations can change their chirality, leading to an imbalance in left- and right-handed quarks (nonzero axial chemical potential $\mu_{5}$); 
$N_{L} - N_{R} = 2n_{f}Q_{w} \propto \mu_{5}$, $n_{f}$ is the number of light quark flavors and $Q_{w}$ is the topological charge of the gluonic configuration. 
Thus, gluonic field configurations with nonzero topological charges induce local parity violation\cite{Kharzeev:1998kz,Kharzeev:2004ey,Kharzeev:2007jp,Fukushima:2008xe,Kharzeev:2015znc}.
It was suggested that in relativistic heavy-ion collisions, where the deconfinement phase transition and 
an extremely strong magnetic field  are present, 
The chirality imbalance of quarks in the local metastable domains will generate an electromagnetic current, $\vec{J} \propto \mu_{5}\vec{B}$, along the direction of the magnetic field.
Quarks hadronize into charged hadrons, leading to an experimentally observable charge separation. 
The measurements of this charge separation provide a means to study the non-trivial QCD topological structures 
in relativistic heavy-ion collisions\cite{Lee:1973iz,Lee:1974ma,Morley:1983wr,Kharzeev:1998kz,Kharzeev:1999cz}.  

In heavy-ion collisions, particle azimuthal angle distribution in momentum space is often described with a Fourier decomposition:
\begin{equation}
	\begin{split}
		\frac{\rm{d}N}{\rm{d}\phi} \propto  1& + 2v_{1}\cos(\Delta\phi) + 2v_{2}\cos(2\Delta\phi) + ...  \\
									  & + 2a_{1}\sin(\Delta\phi) + 2a_{2}\sin(2\Delta\phi) + ...
	\end{split}
	\label{eqThreeCtor3}
\end{equation}
where $\Delta\phi=\phi - \psiRP$, and $\psiRP$ is the reaction-plane direction, 
defined to be the direction of the impact parameter vector and expected to be perpendicular to the magnetic field direction on average. 
The parameters $v_{1}$ and $v_{2}$ account for the directed flow and elliptic flow.
The parameters $a_{1,2}$  can be used to describe the charge separation effects. 
Usually only the first harmonic coefficient $a_{1}$ is considered.
Positively and negatively charged particles have opposite $a_{1}$ values, $a_{1}^{+}=-a_{1}^{-}$, and are proportional to $Q_{w}B$.
However, they average to zero  because of the random topological charge fluctuations from event to event, 
making a direct observation of this parity violation effect impossible.
Indeed, the measured $\mean{a_{1}}$ of both positive and negative charges are less than $5\times10^{-4}$ at the 95\% confidence level
in Au+Au collisions at \sNN = 200 GeV\cite{Adamczyk:2013hsi}. 
The observation of this parity violation effect is possible only via correlations, e.g. measuring 
$\mean{a_{\alpha}a_{\beta}}$ with the average taken over all events in a given event sample.
The $\gamma$ correlator is designed for this propose: 
\begin{equation}
	\begin{split}
		\gamma &= \mean{\cos(\phi_{\alpha}+\phi_{\beta}-2\psi_{RP})} \\  
			   &= \mean{\cos\Delta\phi_{\alpha}\cos\Delta\phi_{\beta}} - \mean{\sin\Delta\phi_{\alpha}\sin\Delta\phi_{\beta}} \\ 
			   &= [\mean{v_{1,\alpha}v_{1,\beta} + B_{in}}]  - [\mean{a_{\alpha}a_{\beta} + B_{out}}]. 
	\end{split}
	\label{eqThreeCtor4}
\end{equation}
$B_{in}$ and $B_{out}$ are the reaction plane dependent backgrounds in in-plane and out-plane directions,
which are assumed to largely cancel out in their difference, 
while there are still residual background contributions (e.g. momentum conservation effect~\cite{Pratt:2010zn,Bzdak:2010fd}). 
At mid-rapidity, the $v_{1}$ is averaged to zero, and the $v_{1}$ contribution ($\mean{v_{1,\alpha}v_{1,\beta}}$) is expected to be small.
Moreover, the $v_{1}$ background is expected to be charge independent. 
By taking the opposite-sign (OS) and same-sign (SS) $\gamma$ difference those charge independent backgrounds can be further cancelled out. 
Thus, usually the $\Delta\gamma$ correlator is used:
\begin{equation}
	\begin{split}
		\Delta\gamma = \gamma_{OS} -\gamma_{SS} 
	\end{split}
	\label{eqThreeCtor5}
\end{equation}
where OS and SS describe the charge sign combinations between the $\alpha$ and $\beta$ particle.

The $\gamma$ correlator can be calculated by the three-particle correlation method without an explicit determination of the reaction plane; 
instead, the role of the reaction plane is played by the third particle, $c$. 
Under the assumption that particle $c$ is correlated  
with particles $\alpha$ and $\beta$ only via common correlation to the reaction plane, we have:

\begin{equation}
	\begin{split}
		\mean{\cos(\phi_{\alpha}+\phi_{\beta}-2\psi_{RP})} = \mean{\cos(\phi_{\alpha}+\phi_{\beta}-2\phi_c)}/v_{2,c} 
	\end{split}
	\label{eqThreeCtor0}
\end{equation}
where $v_{2,c}$ is the elliptic flow parameter of the particle $c$, 
and $\phi_{\alpha}$, $\phi_{\beta}$ and $\phi_{c}$ are the azimuthal angles of particle $\alpha$, $\beta$ and $c$, respectively.

%%%%%%%%%%%%%%%%%%%%%%%%%%%%%%%%%%%%%%%%%%%%%%%%%%%%%%%%%%%%%%%%%%%%%%%%%%%%%%%%%%%%%%%%%%%%%%
\section{Challenges and Strategies}
\begin{figure}[htbp!]
	\centering 
	\includegraphics[width=7.0cm]{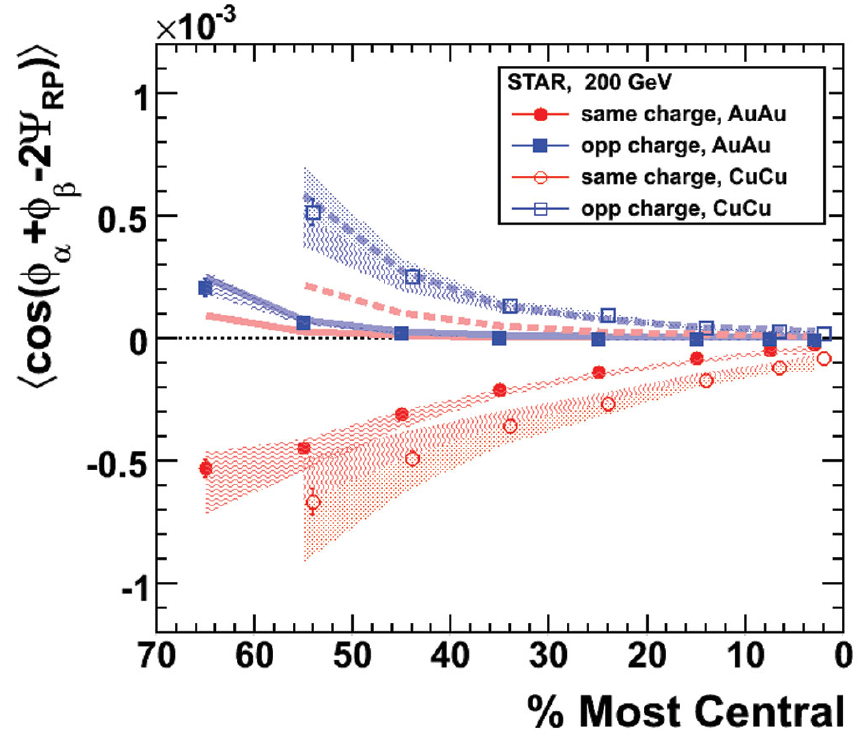} 
	\caption{(Color online)
		$\gamma$ correlator in Au+Au and Cu+Cu collisions at \sNN = 200 GeV. Shaded bands represent uncertainty from the measurement of $v_{2}$. 
		The thick solid (Au+Au) and dashed (Cu+Cu) lines represent HIJING calculations of the contributions from three-particle
		correlations. Collision centrality increases from left to right\cite{Abelev:2009ad}.
	}   
	\label{FG_FirstCME}
\end{figure}

\begin{figure}[htbp!]
	\centering 
	\includegraphics[width=13.0cm]{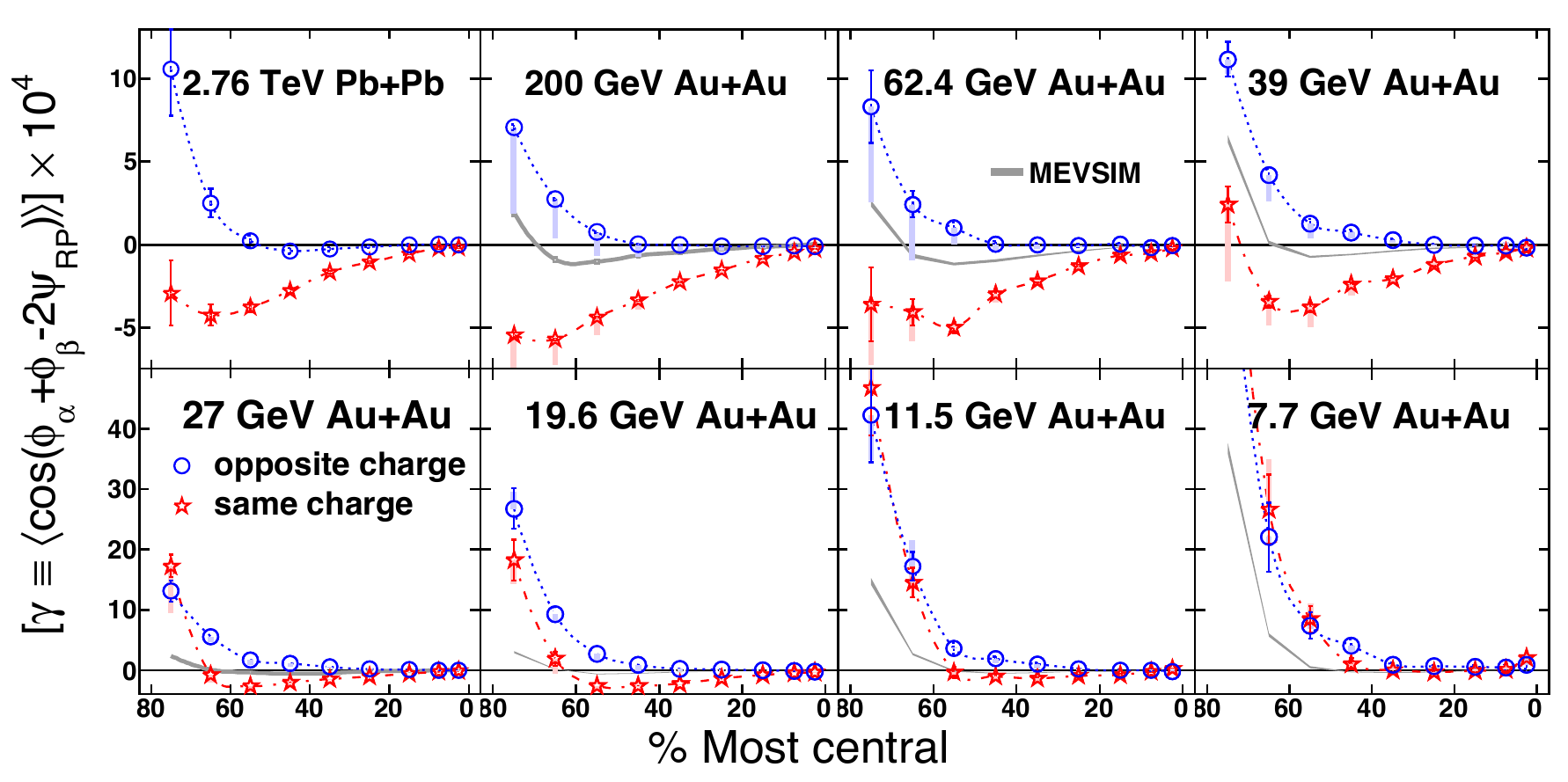} 
	\caption{(Color online)
		$\gamma$ correlator as a function of centrality for Au+Au collisions at 7.7-200 GeV\cite{Abelev:2009ac,Abelev:2009ad,Adamczyk:2014mzf}, and for Pb+Pb collisions
		at 2.76 TeV\cite{Abelev:2012pa}. Grey curves are the charge independent results from MEVSIM calculations\cite{Adamczyk:2014mzf}. 
	}   
	\label{FG_BESCME}
\end{figure}

A significant $\dg$ has indeed been observed in heavy-ion collisions at RHIC and LHC\cite{Abelev:2009ad,Abelev:2009ac,Adamczyk:2013hsi,Adamczyk:2013kcb,Adamczyk:2014mzf,Abelev:2012pa,Ajitanand:2010}. 
The first $\gamma$ measurement was made by the STAR collaboration at RHIC in 2009~\cite{Abelev:2009ad}.
Fig.~\ref{FG_FirstCME} shows their $\gamma$ correlator as a function of the collision centrality in Au+Au and Cu+Cu collisions at \sNN = 200 GeV.
Charge dependent signal of the same-sign and opposite-sign charge $\gamma$ correlators have been observed.
Similarly, Fig.~\ref{FG_BESCME} shows the $\gOS$ and $\gSS$ correlator as a function of the collision centrality in Au+Au collisions at \sNN = 7.7-200 GeV from STAR\cite{Adamczyk:2014mzf} and in Pb+Pb collisions at 2.76 TeV from ALICE\cite{Abelev:2012pa}.
At high collision energies, charge dependent signals are observed, and $\gOS$ is larger than $\gSS$. 
The difference between $\gOS$ and $\gSS$, $\dg$, decreases with increasing centrality, 
which would be consistent with expectation of the magnetic field strength to decrease with increasing centrality. 
At the low collision energy of \sNN =7.7 GeV, the difference between the $\gOS$ and $\gSS$ disappears, 
which could be consistent with the disappearance of the CME in the hadronic dominant stage at this energy. 
Thus, these results are qualitatively consistent with the CME expectation.

There are, however, mundane physics that could generate the same effect as the CME in the $\dg$ variable,
which contribute to the background in the $\dg$ measurements.
An example is the resonance or cluster decay (coupled with $v_{2}$) background\cite{Wang:2009kd,Wang:2016iov,Voloshin:2004vk};
the $\dg$ variable is ambiguous between a back-to-back OS pair from the CME 
perpendicular to $\psi_2$ and an OS pair from a resonance decay along $\psi_2$. 
Calculations with local charge conservation and momentum conservation effects can almost fully account for the measured $\dg$ signal at RHIC\cite{Pratt:2010zn,Schlichting:2010qia,Bzdak:2010fd}. 
A Multi-Phase Transport (AMPT)\cite{Zhang:1999bd,Lin:2001zk,Lin:2004en} model simulations can also largely account for the measured $\dg$ signal\cite{Ma:2011uma,Shou:2014zsa}.
In general, these backgrounds are generated by two particle correlations coupled with elliptic flow ($v_{2}$):
\begin{equation}
	\begin{split}
		\mean{\cos(\phi_{\alpha}+\phi_{\beta}-2\psi_{RP})} &= \mean{\cos(\phi_{\alpha}+\phi_{\beta}-2\phi_{reso.} + 2\phi_{reso.} -2\psi_{RP})}, \\
														   &\approx \mean{\cos(\phi_{\alpha}+\phi_{\beta}-2\phi_{reso.}}\times v_{2,reso.}. 
	\end{split}
	\label{EQ_2}
\end{equation}
Thus, a two particle correlation of $\mean{\cos(\phi_{\alpha}+\phi_{\beta}-2\phi_{reso.}}$ from resonance (cluster) decays, 
coupled with the $v_{2}$ of the resonance (cluster), will lead to a $\dg$ signal.

Experimentally, various proposals and attempts have been put forward to reduce or eliminate backgrounds, 
exploiting their dependences on $v_{2}$ and two particle correlations. 
(1) Using the event shape selection, by varying the event-by-event $v_{2}$ exploiting statistical (event-by-event $v_{2}, q_{2}$ methods)\cite{Adamczyk:2013kcb,Wen:2016zic} and dynamical fluctuations (event shape engineering method)\cite{Acharya:2017fau,Sirunyan:2017quh},
it is expected that the $v_{2}$ independent contribution to the $\dg$ can be extracted. 
(2) Isobaric collisions and Uranium+Uranium collisions have been proposed\cite{Voloshin:2010ut} to take advantage of the different nuclear properties (such as proton number, shape). 
(3) Control experiments of small system p+A or d+A collisions are used to study the background behavior\cite{Khachatryan:2016got,Zhao:2017wck}, 
where backgrounds and possible CME signals are expected to be uncorrelated because the participant plane\cite{Alver:2008zza} and the magnetic field direction are uncorrelated due to geometry fluctuations in these small system collisions. 
(4) A new idea of differential measurements with respect to reaction plane and participant plane are proposed\cite{Xu:2017qfs,Xu:2017zcn}, 
which takes advantage of the geometry fluctuation effects on the participant plane and the magnetic field direction in A+A collisions. 
%(5) A new method exploiting the invariant mass dependence of the $\dg$ measurements is devised to provide the identification capability of the resonance decay backgrounds, 
%hence to remove those backgrounds and enhance the sensitivity to CME signal\cite{Zhao:2017nfq,Zhao:2017wck}.
(5) A new method exploiting the invariant mass dependence of the $\dg$ measurements is devised, which identifies and removes the resonance decay backgrounds, to enhance the sensitivity of CME measurement\cite{Zhao:2017nfq,Zhao:2017wck}.
(6) New $R(\Delta S)$ correlator is designed to detect the CME-driven charge separation\cite{Magdy:2017yje,Ajitanand:2010rc}. 
In the following sections we will review these proposals and attempts in more detail.

%%%%%%%%%%%%%%%%%%%%%%%%%%%%%%%%%%%%%%%%%%%%%%%%%%%%%%%%%%%%%%%%%%%%%%%%%%%%%%%%%%%%%%%%%%%%%%
\section{Event-by-event selection methods}
The main background sources of the $\dg$ measurements are from the elliptic flow ($v_{2}$) induced effects.
These backgrounds are expected to be proportional to the $v_{2}$. 
One possible way to eliminate or suppress these $v_{2}$ induced backgrounds is to select ``spherical'' events with $v_{2}=0$ 
exploiting the statistical and dynamical fluctuations of the event-by-event $v_{2}$.
Due to finite multiplicity fluctuations,
one can easily vary the shape of the final particle momentum space, 
which is directly related to the $v_{2}$ backgrounds\cite{Adamczyk:2013kcb}. 

\begin{figure}[htbp!]
	\centering 
	\includegraphics[width=6.2cm]{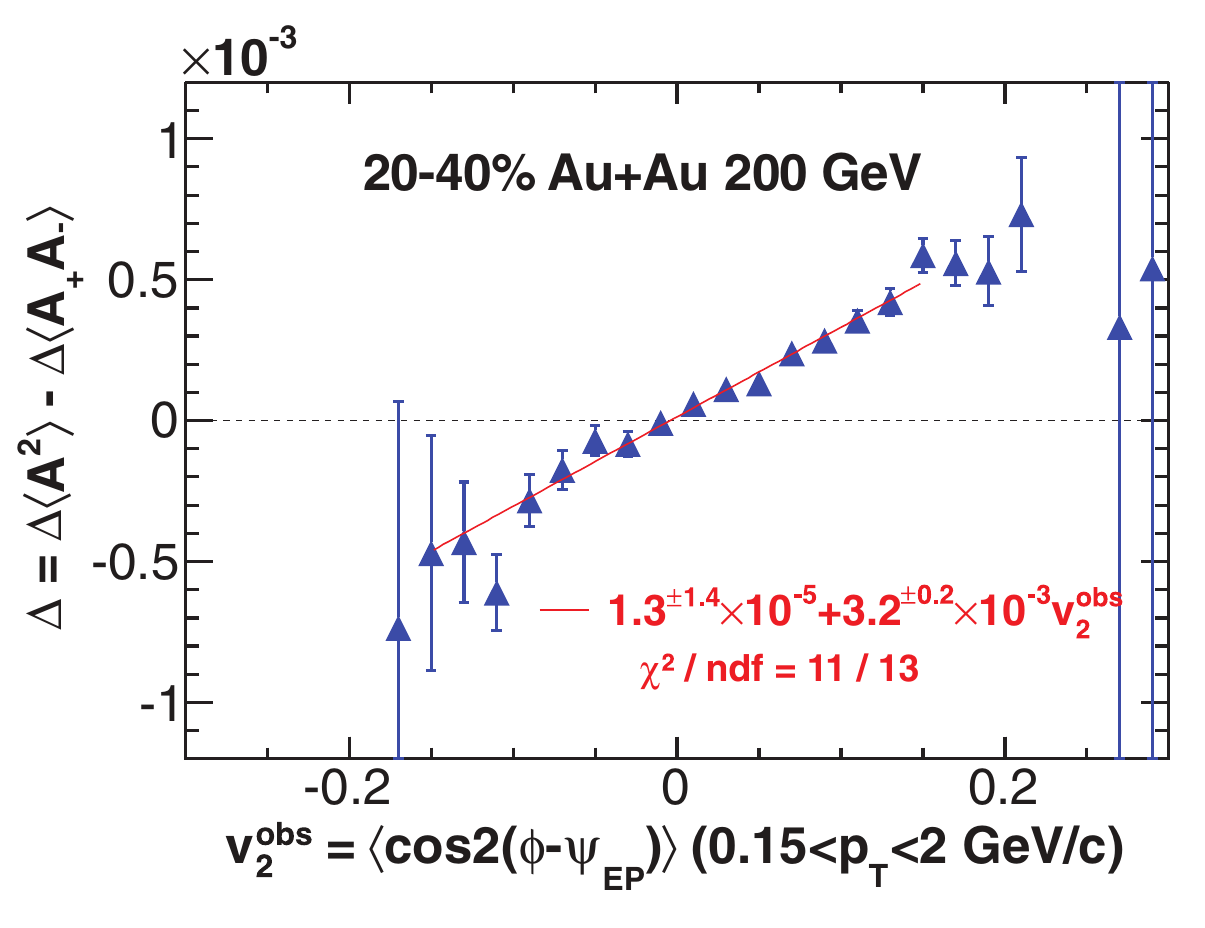} 
	\caption{(Color online)
		charge multiplicity asymmetry correlations ($\Delta$) as a function of $\vobs$ from Au+Au collisions at \sNN = 200 GeV\cite{Adamczyk:2013kcb}. 
	}   
	\label{FG_STARese}
\end{figure}

By using the event-by-event $v_{2}$, STAR has carried out the first attempt to remove the backgrounds\cite{Adamczyk:2013kcb}.
Fig.~\ref{FG_STARese} shows the charge multiplicity asymmetry correlator ($\Delta$) as a function of the event-by-event $v_{2}$.
The event-by-event $v_{2}$ ($\vobs$) can be measured by the $Q$ vector method:
\begin{equation}
	\begin{split}
		&\vobs = Q^{*}\it{q}_{EP},   \\
		&Q=\frac{1}{N}\sum_{j=1}^{N} e^{2i\phi_j}, \ \it{q}_{EP}=e^{2i\psiEP}, \\
	\end{split}
	\label{EQ_ESE1}
\end{equation}
where $Q$ sums over particles (used for the $\Delta$ correlator) in each event; $\psiEP$ is the event plane (EP) azimuthal angle,
reconstructed from final-state particles, as a proxy for participant plane ($\psiPP$) that is not experimentally accessible.
To avoid self-correlation, particles used for the EP calculations are exclusive to the particles used for $Q$ and $\Delta$ correlator.  
The results show strong correlation between the $\Delta$ correlator and the $\vobs$.
By selecting the events with $\vobs=0$, the $\Delta$ correlator is largely reduced\cite{Adamczyk:2013kcb,Tu:2015qm}. 
The $\dg$ correlator shows similar correlation with $\vobs$ from the preliminary STAR data\cite{Zhao:2017ckp}.  

A similar method selecting events with the $q_{2}$ (see Eq.~\ref{EQ_ESE2}) variable has been proposed recently\cite{Wen:2016zic}. 
To suppress the $v_{2}$ related background, a tighter cut, $q_{2} = 0$, is proposed to extract signal.
The cut is tighter because $q_{2} = 0$ corresponds to a zero $2^{nd}$ harmonic to any plane, 
while $\vobs = 0$ corresponds to zero $2^{nd}$ harmonic with respect to the reconstructed EP in the event. 

These methods assume the background to be linear in $v_{2}$ of the final-state particles.  
However, the background arises from the correlated pairs (resonance/cluster decay) coupled with the $v_{2}$ of the parent sources, 
not the final-state particles.
In case of resonance decays: $\dg = \mean{\cos(\phi_{\alpha}+\phi_{\beta}-2\psi_{RP})} = \mean{\cos(\phi_{\alpha}+\phi_{\beta}-2\phi_{\rm reso.}}v_{2,\rm reso.}$,  
where $\mean{\cos(\phi_{\alpha}+\phi_{\beta}-2\phi_{\rm reso.}}$ depends on the resonance decay kinematics, 
and $v_{2,\rm{reso.}}$ is the $v_{2}$ of the resonances, not the decay particles'. 
It is difficult, if not at all impossible, to ensure the $v_{2}$ of all the background sources to be zero.
Thus, it is challenging to completely remove flow background by using the event-by-event $v_{2}$ or $q_{2}$ methods\cite{Wang:2016iov}.

\section{Event shape engineering}
Because of dynamical fluctuations of the event-by-event $v_{2}$, 
one could possibly select events with different initial participant geometries (participant eccentricities) 
even with the same impact parameter\cite{Voloshin:2010ut,Bzdak:2011np,Schukraft:2012ah}.
By restricting to a narrow centrality, 
while varying event-by-event $v_{2}$, one is presumably still fixing the magnetic filed (mainly determined by the initial distribution of the spectator protons)\cite{Bzdak:2011np}.
This provides a way to decouple the magnetic field and the $v_{2}$,
and thus a possible way to disentangle background contributions from potential CME signals. 
This is usually called the event shape engineering (ESE) method\cite{Schukraft:2012ah}.

In ESE, instead of selecting on $\vobs$, 
one use the flow vector to possibly access the initial participant geometry,
selecting different event shapes by making use of the dynamical fluctuations of $v_{2}$\cite{Schukraft:2012ah,Voloshin:2010ut,Bzdak:2011np}.
The ESE method is performed based on the magnitude of the second-order reduced flow vector, $q_{2}$\cite{Adler:2002pu}, defined as:
\begin{equation}
	\begin{split}
		&q_{2}   = \frac{|Q_{2}|}{\sqrt{M}}, \\
		&|Q_{2}| = \sqrt{Q_{2,x}^{2}+Q_{2,y}^{2}}, \\
		&Q_{2,x} = \sum_{i}w_{i}\cos(2\phi_{i}), \ Q_{2,y} = \sum_{i}w_{i}\sin(2\phi_{i}), \\	
	\end{split}
	\label{EQ_ESE2}
\end{equation}
where $|Q_{2}|$ is the magnitude of the second order harmonic flow vector and M is the multiplicity.
The sum runs over all particles/hits, $\phi_{i}$ is the azimuthal angle of the $i$-th particle/hit, and $w_{i}$ is the weight 
(usually taken to be the $\pt$ of the particle or energy deposition of the hit)\cite{Acharya:2017fau,Sirunyan:2017quh}.

\begin{figure}[htbp!]
	\centering 
	\includegraphics[width=10cm]{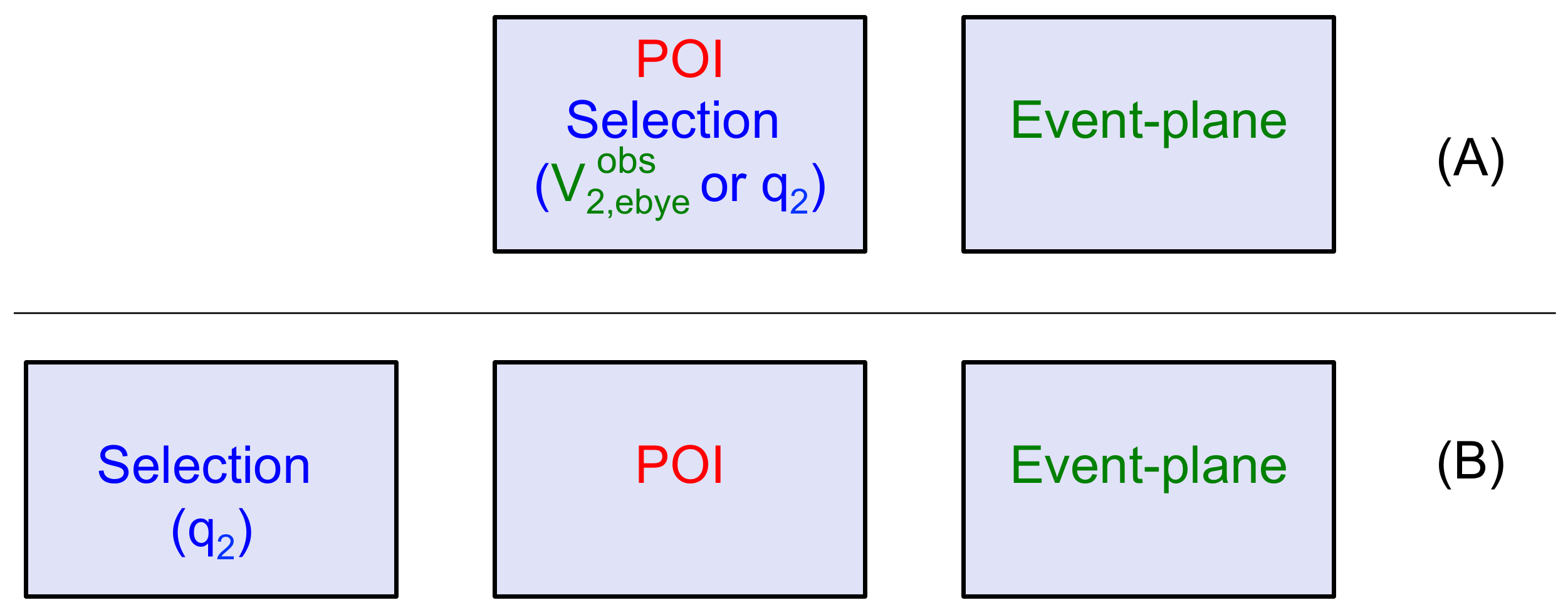} 
	\caption{(Color online)
		Schematic comparison of the event-by-event selection (A) and the ESE (B) methods.
		Different boxes represent different phase spaces, usually displaced in $\eta$.
		The green color for $\vobs$ in (A) reflects that the $\vobs$ is calculated with respect to the event-plane.  
		The $\dg$ correlator is usually calculated from the correlation between the particle of interest 
		(POI, here the POI refers to the $\alpha$ and $\beta$ particles in Eq.~\ref{eqThreeCtor0}) and the event-plane. 
	}   
	\label{FG_ESEcom}
\end{figure}

Figure~\ref{FG_ESEcom} is a schematic comparison of the event-by-event selection and the ESE methods. 
Basically, the most important difference between these two groups of methods lies in which phase space to calculate the $\vobs$ or $q_{2}$ variables for event selection. 
In the event-by-event selection methods, the same phase space of the particle of interest (POI) is used for event selections, 
thus these methods take advantage of statistical as well as dynamical fluctuations of the POI.
In the ESE method, a different phase space is used (often displaced in $\eta$), so that the event selection is dominated by the dynamical fluctuations, 
because statistical fluctuations of POI and event selection are independent.
The dynamical fluctuations stem out of the common origin of the initial participant geometries.
Thus a zero $q_{2}$ should correspond to an average zero $v_{2}$ of the background sources of the POI. 
However, a zero $q_{2}$ is unlikely accessible directly from data, so extrapolation is often involved.

\begin{figure}[htbp!]
	\centering 
	\includegraphics[width=6.2cm]{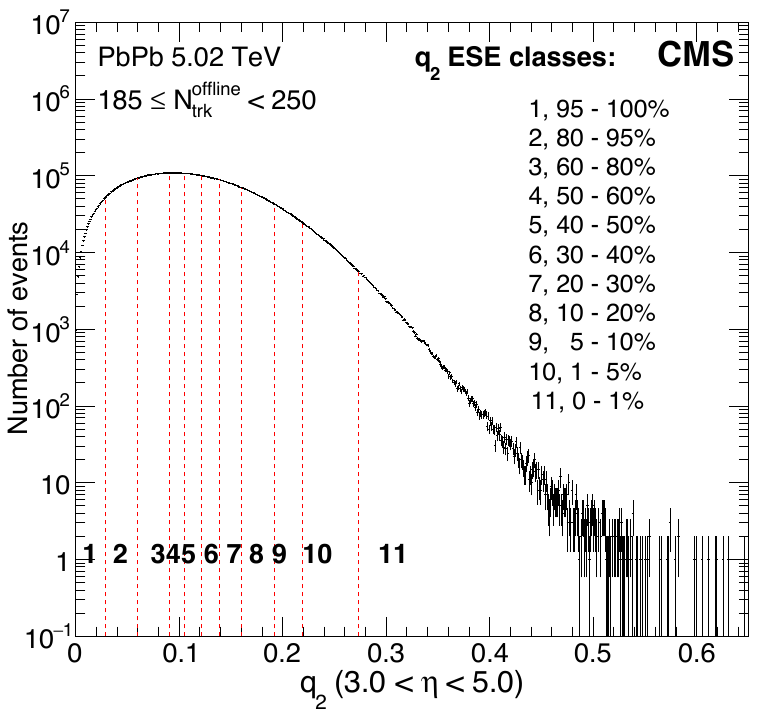} 
	\includegraphics[width=6.2cm]{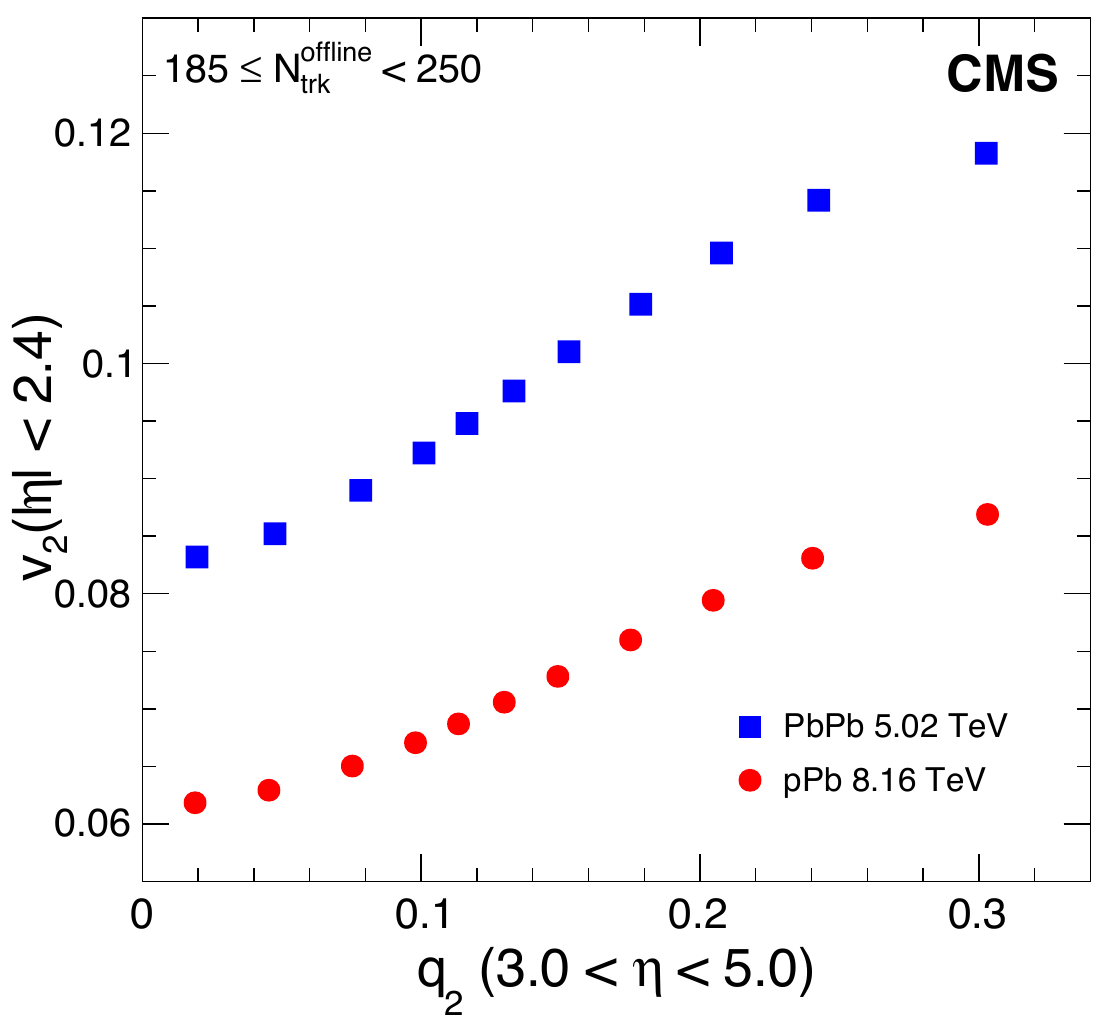} 
	\caption{(Color online)
		(Left) The $q_{2}$ distributions in multiplicity range $185 \leq N^{\rm offline}_{\rm trk} < 250$ in Pb+Pb collisions.
		Red dash line represents the selection used to divide the events into multiple $q_{2}$ classes.
		(Right) The correlation between $v_{2}$ and $q_{2}$ in p+Pb and Pb+Pb collisions based on the $q_{2}$ selections of the events\cite{Sirunyan:2017quh}. 
	}   
	\label{FG_CMSeseA}
\end{figure}

Figure~\ref{FG_CMSeseA}(left) shows the $q_{2}$ distribution in Pb+Pb collisions from CMS\cite{Sirunyan:2017quh}. 
The events of a narrow multiplicity bin are divided into several classes with each corresponding to a fraction of the full distribution, 
where 0-1\% represents the events with the largest $q_{2}$ value, and 95-100\% corresponds to the events with smallest $q_{2}$ value, and so on. 
Fig.~\ref{FG_CMSeseA}(right) shows that the $v_{2}$ is closely proportional to $q_{2}$, 
suggesting those two quantities are strongly correlated because of the common initial-state geometry\cite{Sirunyan:2017quh}.
One could thus use the $q_{2}$ to select events with different $v_{2}$, and study the $v_{2}$ dependence of the $\dg$ correlator. 
In a similar way, the $\dg$ correlator is also calculated in each $q_{2}$ class.

\begin{figure}[htbp!]
	\centering 
	\includegraphics[width=6.2cm]{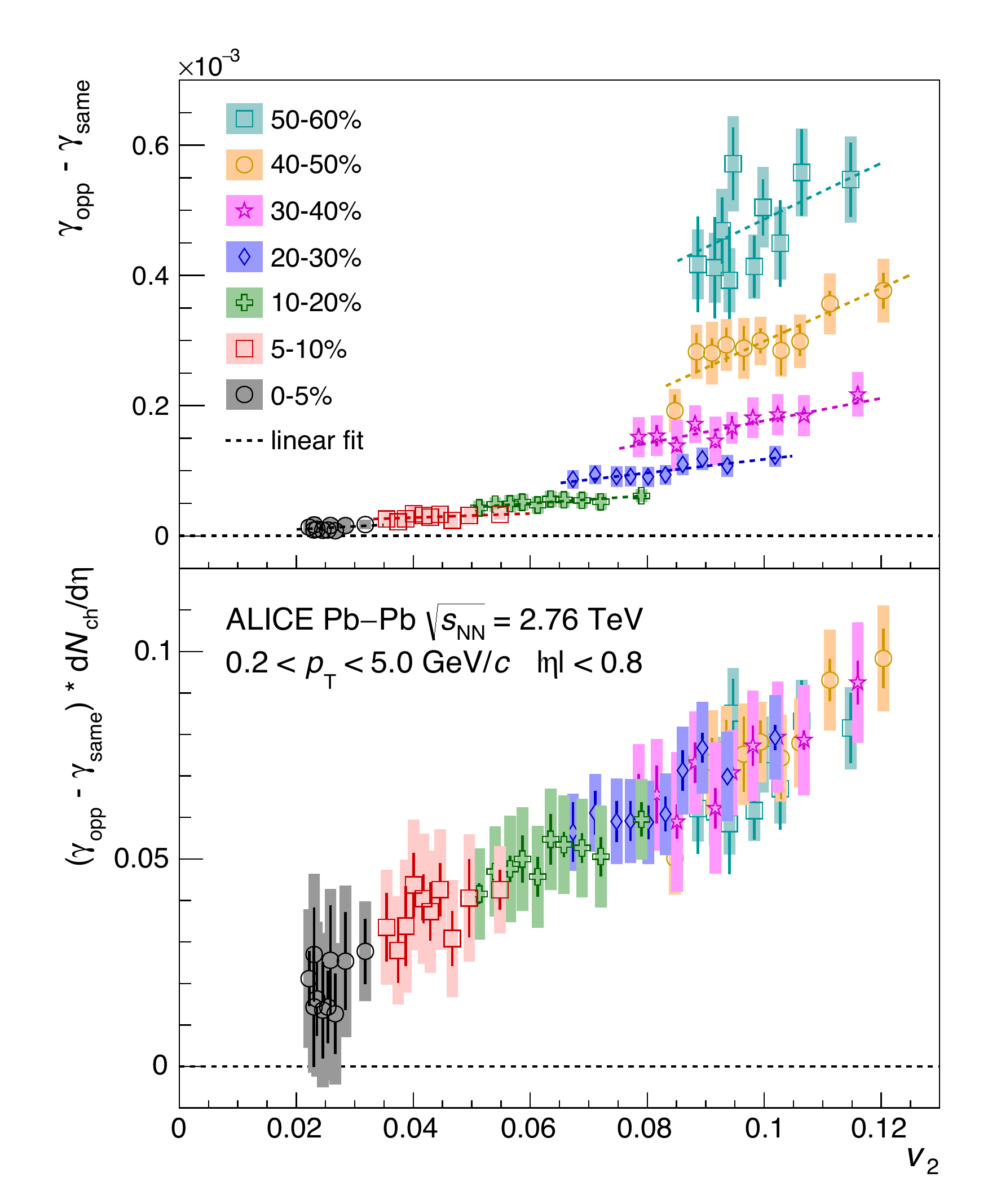} 
	\includegraphics[width=6.2cm]{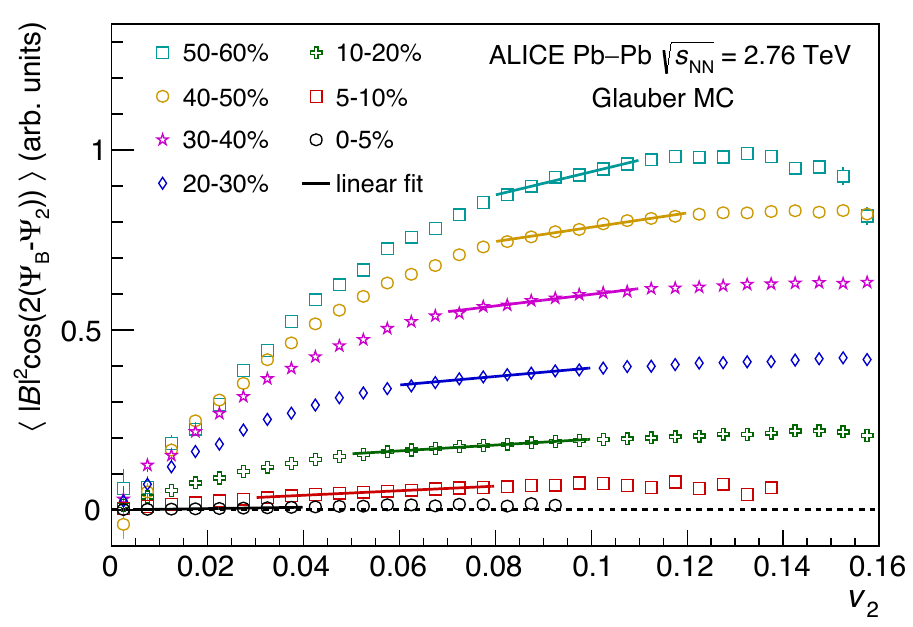} 
	\caption{(Color online)
		(Left top) The $\dg$ correlator and 
		(Left bottom) the charged-particle density scaled correlator $\dg \cdot dN_{ch}/d\eta$ 
		as a function of $v_{2}$ for shape selected events with $q_{2}$ for various centrality classes.
		Error bars (shaded boxes) represent the statistical (systematic) uncertainties.
		(Right) The expected dependence of the CME signal on $v_{2}$ for various centrality classes from a MC-Glauber simulation\cite{Miller:2007ri}. 
		The solid lines depict linear fits based on the $v_{2}$ variation observed within each centrality interval\cite{Acharya:2017fau}.
	}   
	\label{FG_ALICEeseB}
\end{figure}

Fig.~\ref{FG_ALICEeseB}(upper left) shows $\dg$ correlator as a function of $v_{2}$ in different centralities in Pb+Pb collisions from ALICE\cite{Acharya:2017fau}. 
To compensate for the dilution effect, $\dg$ correlator was multiplied by the charged-particle density in a given centrality bin ($dN_{ch}/d\eta$) in the lower left panel.
The results show strong dependence on $v_{2}$, and the $dN_{ch}/d\eta$ scaled $\dg$ correlator falls approximately onto the same linear trend for different centralities.
This is qualitatively consistent with the expectation from background effects, such as resonance decay coupled with $v_{2}$\cite{Hori:2012kp,Wang:2009kd,Wang:2016iov}. 
Therefore, the observed dependence on $v_{2}$ indicates a large background contribution to $\dg$ correlator\cite{Acharya:2017fau}.

By restricting to a given narrow centrality, the event shape selection is expected to be less affected by the magnetic field\cite{Bzdak:2011np}. 
The different dependences of the CME signal and background on $v_{2}$ ($q_{2}$) could possibly be used to disentangle the CME signal from background. 
Fig.~\ref{FG_ALICEeseB}(right) shows the $v_{2}$ dependence of the $\mean{|\textbf B|^{2} \cos2(\psi_{B} - \psi_{2})}$ 
from Monte Carlo Glauber calculation\cite{Acharya:2017fau}. 
The CME signal is assumed to be proportional to $\mean{|\textbf B|^{2} \cos2(\psi_{B} - \psi_{2})}$, 
where $|\textbf B|$ and $\psi_{B}$ are the magnitude and azimuthal direction of the magnetic field.
The calculation shows that the CME signal weakly depends on $v_{2}$ within each given centrality (Fig.~\ref{FG_ALICEeseB} right panel) and approximately linear.
To extract the contribution of the possible CME signal to the current $\dg$ measurements, 
a linear function is fit to the data:
\begin{equation}
	\begin{split}
		F_{1}(v_{2}) = p_{0}(1+p_{1}(v_{2}-\mean{v_{2}})/\mean{v_{2}}). 
	\end{split}
	\label{EQ_ESE3}
\end{equation}
Here $p_{0}$ accounts for a overall scale. $p_{1}$ is the normalised slope, 
reflecting the dependence on $v_{2}$. 
In a pure background scenario, the $\dg$ correlator is linearly proportional to $v_{2}$ and the $p_{1}$ parameter is equal to unity, 
Eq.~\ref{EQ_ESE3} is reduced to $F_{1}(v_{2}) = p_{0}v_{2}/\mean{v_{2}} \propto v_{2} $. 
On the other hand, a significant CME contribution would result in non-zero intercepts at $v_{2}$ = 0 of the linear functional fits shown in Fig.~\ref{FG_ALICEeseB}(top left). 

\begin{figure}[htbp!]
	\centering 
	\includegraphics[width=6.2cm]{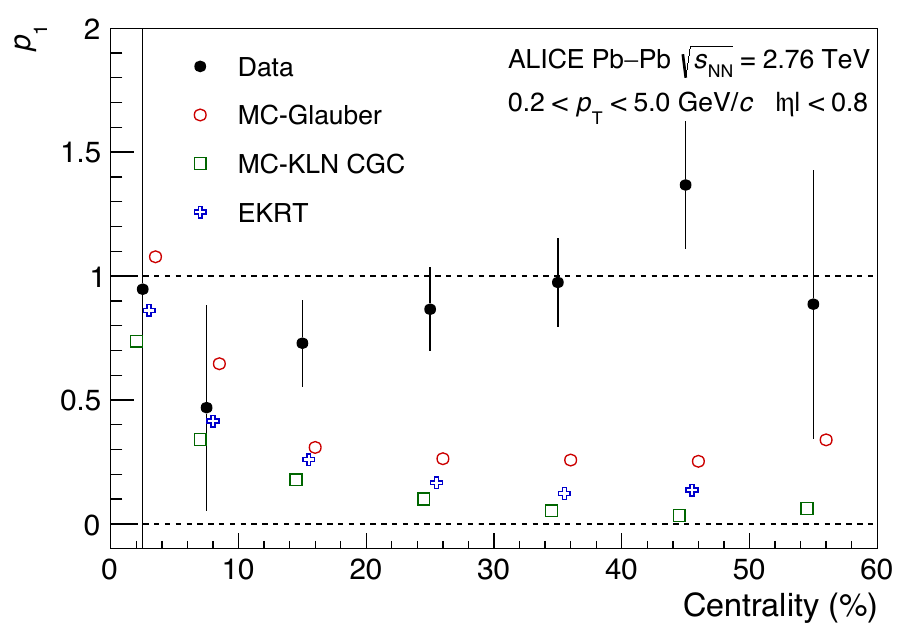} 
	\includegraphics[width=6.2cm]{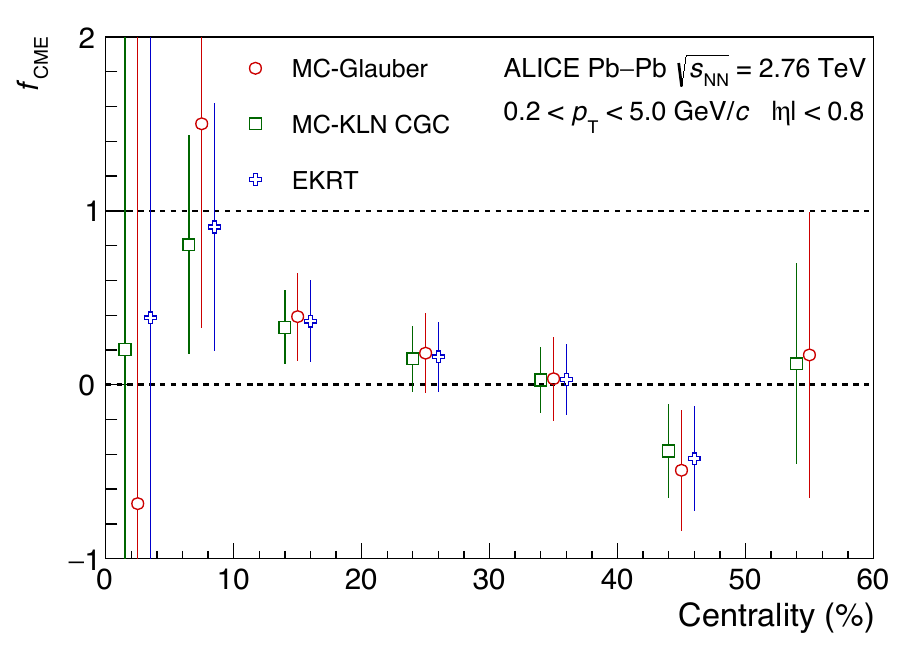} 
	\caption{(Color online)
		(Left) Centrality dependence of the $p_{1}$ parameter from a linear fit to the $\dg$ correlator in Pb+Pb collisions from ALICE 
		and from linear fits to the CME signal expectations from MC-Glauber\cite{Miller:2007ri}, 
		MC-KLN CGC\cite{Drescher:2007ax,ALbacete:2010ad} and EKRT\cite{Niemi:2015qia} models.
		(Right) Centrality dependence of the CME fraction extracted from the slope parameter of fits to data and different models. 
		Points from MC simulations are slightly shifted along the horizontal axis for better visibility. Only statistical uncertainties are shown\cite{Acharya:2017fau}.
	}   
	\label{FG_ALICEeseC}
\end{figure}

In a naive two components model with signal and background, a measured observable ($O_{m}$) can be expressed as: 
\begin{equation}
	\begin{split}
		\frac{S}{S+B}\times O_{S} + \frac{B}{S+B}\times O_{B} = O_{m}, \\  
	\end{split}
	\label{EQ_ESE4}
\end{equation}
$O_{S}$ and $O_{B}$ are the values of the observable $O$ from signal and background, $\frac{S}{S+B}$ represents the fraction of signal contribution in the measurement.
The $p_{1}$ from the fit to the measured data is thus a combination of CME signal slope ($p_{1,Sig} = p_{1,MC}$) and the background slope ($ p_{1,Bkg} \equiv 1$): 
\begin{equation}
	\begin{split}
		&f_{CME}\times p_{1,Sig} + (1-f_{CME})\times p_{1,Bkg} = p_{1,data},	
	\end{split}
	\label{EQ_ESE5}
\end{equation}
where $f_{CME} = \frac{\dg_{CME}}{\dg_{CME}+\dg_{Bkg}}$ represents the CME fraction to the $\dg$ correlator from the measurements, 
and $p_{1,MC}$ is the slope parameter from the MC calculations in Fig.~\ref{FG_ALICEeseB} right panel.
Figure~\ref{FG_ALICEeseC}(left) shows the centrality dependence of $p_{1,data}$ from fits to data and $p_{1,MC}$  
to the signal expectations based on MC-Glauber, MC-KLN CGC and EKRT models\cite{Acharya:2017fau}. 
Fig.~\ref{FG_ALICEeseC}(right) presents the estimated $f_{CME}$ from the three models. 
The $f_{CME}$ extracted for central (0-10\%) and peripheral (50-60\%) collisions have currently large uncertainties.
Combining the points from 10-50\% neglecting a possible centrality dependence gives $f_{CME} = 0.10 \pm 0.13$, $f_{CME} = 0.08 \pm 0.10$ and $f_{CME} = 0.08 \pm 0.11$ 
for the MC-Glauber, MC-KLN CGC and EKRT models inputs of $p_{1,MC}$, respectively. 
These results are consistent with zero CME fraction and correspond to upper limits on $f_{CME}$ of 33\%, 26\% and 29\%, respectively, 
at 95\% confidence level for the 10-50\% centrality interval\cite{Acharya:2017fau}. 

The above analysis is model-dependent, relying on precise modeling of the magnetic field with a given centrality. The CMS collaboration took a different approach, cutting on very narrow centrality bins and assuming the magnetic field to be constant within each centrality bin\cite{Acharya:2017fau}.
The background contribution to the $\gamma$ correlator is approximated to be\cite{Bzdak:2012ia}: 
\begin{equation}
	\begin{split}
		\gamma & \equiv \mean{\cos(\phi_{\alpha}-\phi_{\beta}-2\psi_{RP})},  \\ 
		\gamma^{bkg} & =\kappa_{2}\mean{\cos(\phi_{\alpha}-\phi_{\beta})}\mean{\cos2(\phi_{\beta}-\psi_{RP})}   \\
		    & =\kappa_{2} \delta v_{2},  \\	
		\delta &\equiv \mean{\cos(\phi_{\alpha}-\phi_{\beta})}. \\
	\end{split}
	\label{EQ_ESE6}
\end{equation}
Here, $\delta$ represents the charge-dependent two-particle azimuthal correlator and $\kappa_{2}$ is a constant parameter, independent of $v_{2}$, 
but mainly determined by the kinematics and acceptance of particle detection\cite{Bzdak:2012ia}.
The $\delta$, $\gamma$ and $v_{2}$ are experimental measured observables. 
With event shape engineering to select event with different $v_{2}$, the above Eq.\ref{EQ_ESE6} can be tested.
The charge-independent background sources are eliminated by taking the difference of the correlators ($\gamma, \delta$) between same- and opposite-sign pairs.
In the background scenario, the $\dg$ is expected to be:
\begin{equation}
	\begin{split}
		\dg =\kappa_{2} \Delta\delta v_{2}.  \\ 
	\end{split}
	\label{EQ_ESE7}
\end{equation}
A linear function was used to extract the $v_{2}$-independent fraction of the $\dg$ correlator:
\begin{equation}
	\begin{split}
		\dg/\Delta\delta = a_{norm} v_{2} + b_{norm},  \\ 
	\end{split}
	\label{EQ_ESE8}
\end{equation}
where $b_{norm}$ could be possibly the contribution from CME signal.

\begin{figure}[htbp!]
	\centering 
	\includegraphics[width=6.2cm]{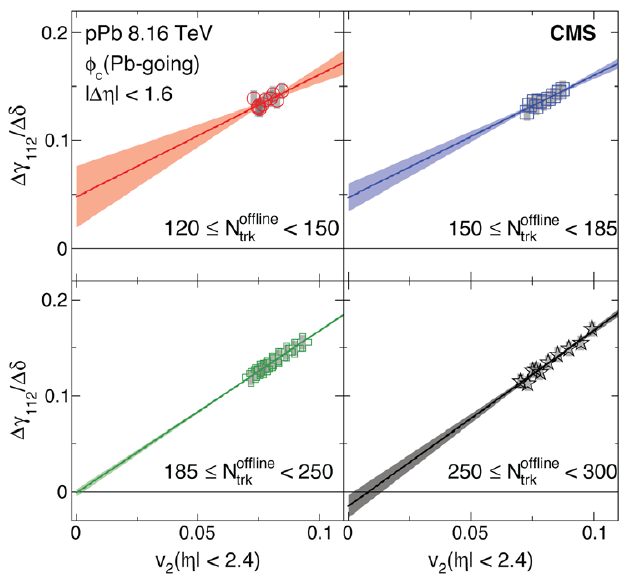} 
	\includegraphics[width=6.2cm]{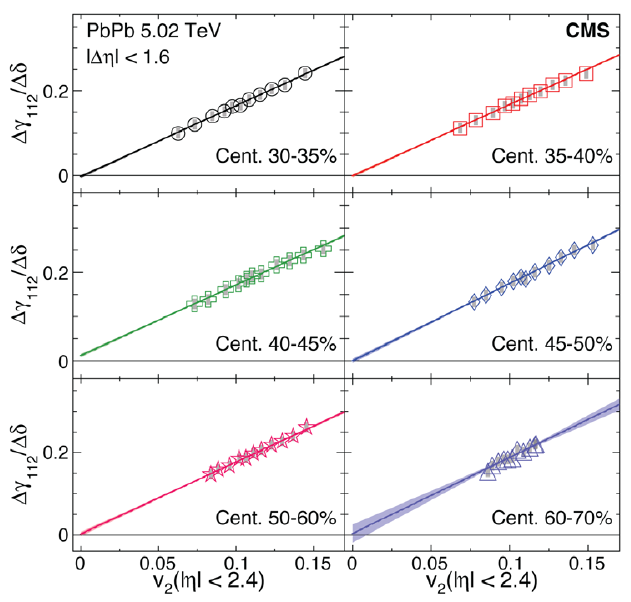} 
	\caption{(Color online)
		The ratio between $\dg$ ($\dg_{112}$) and $\Delta\delta$ correlators, $\dg/\Delta\delta$, 
		averaged over $|\Delta\eta|< 1.6$ as a function of
		$v_{2}$ evaluated in each $q_{2}$ class, for different multiplicity ranges in p+Pb (left) and Pb+Pb (right) collisions\cite{Sirunyan:2017quh}. 
	}   
	\label{FG_CMSESEA}
\end{figure}

\begin{figure}[htbp!]
	\centering 
	\includegraphics[width=6.2cm]{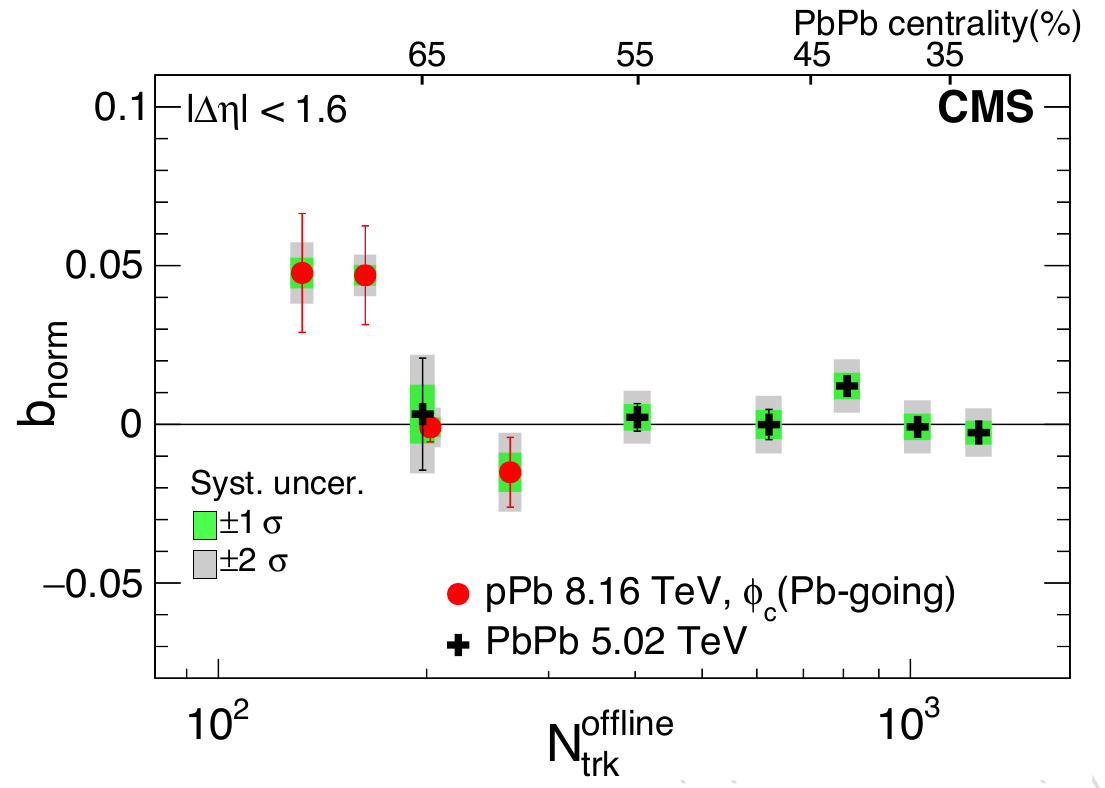} 
	\includegraphics[width=6.2cm]{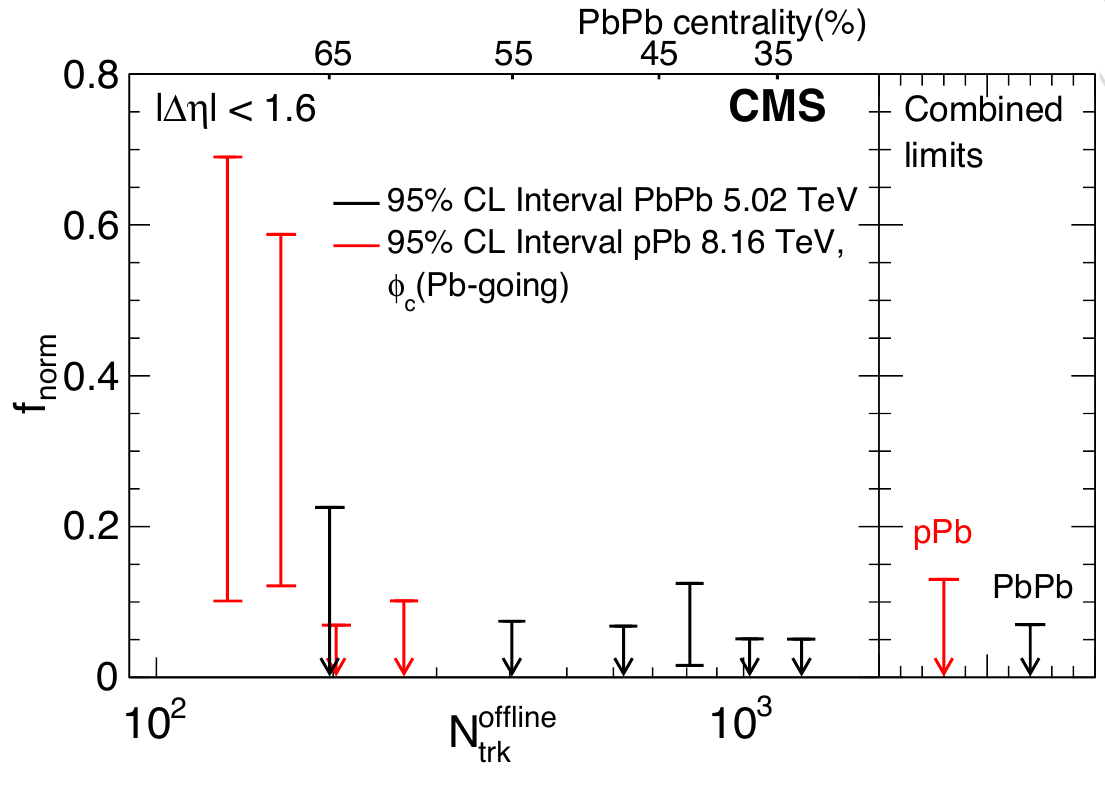} 
	\caption{(Color online)
		(Left) Extracted intercept parameter $b_{norm}$ and (Right) corresponding upper limit of the 
		fraction of $v_{2}$-independent $\dg$ correlator component, averaged over $|\Delta\eta|< 1.6$, as a
		function of $N^{\rm offline}_{\rm trk}$ in p+Pb and Pb+Pb collisions from CMS\cite{Sirunyan:2017quh}. 
	}   
	\label{FG_CMSESEB}
\end{figure}

Figure~\ref{FG_CMSESEA} shows the ratio of $\Delta\gamma/\Delta\delta$ as function of $v_{2}$ for different multiplicity ranges
in p+Pb (left) and Pb+Pb (right) collisions\cite{Sirunyan:2017quh}. 
The values of the intercept parameter $b_{norm}$ are shown as a function of event multiplicity in Fig.~\ref{FG_CMSESEB}(left).
Within statistical and systematic uncertainties, no significant positive value for $b_{norm}$ is observed. 
Result suggests that the $v_{2}$-independent contribution to the $\dg$ correlator is consistent with zero, 
and the $\dg$ results are consistent with the background-only scenario of charge-dependent two-particle correlations\cite{Sirunyan:2017quh}.
Based on the assumption of a nonnegative CME signal, the upper limit of the $v_{2}$-independent fraction in the $\dg$ correlator is obtained from the Feldman-Cousins approach\cite{Feldman:1997qc} with the measured statistical and systematic uncertainties. 
Fig.~\ref{FG_CMSESEB}(right) shows the upper limit of the fraction $f_{norm}$, 
the ratio of the $b_{norm}$ value to the value of $\mean{\dg}/\mean{\Delta\delta}$, at 95\% CL as a function of event multiplicity. 
The $v_{2}$-independent component of the $\dg$ correlator is less than 8-15\% for most of the multiplicity or centrality range. 
The combined limits from all presented multiplicities and centralities are also shown in p+Pb and Pb+Pb collisions. 
An upper limit on the $v_{2}$-independent fraction of the three-particle correlator, or possibly the CME signal contribution, 
is estimated to be 13\% in p+Pb and 7\% in Pb+Pb collisions, at 95\% CL. 
The results are consistent with a $v_{2}$-dependent background-only scenario, 
posing a significant challenge to the search for the CME in heavy ion collisions using three-particle azimuthal correlations\cite{Sirunyan:2017quh}.

\begin{figure}[htbp!]
	\centering 
	\includegraphics[width=6.2cm]{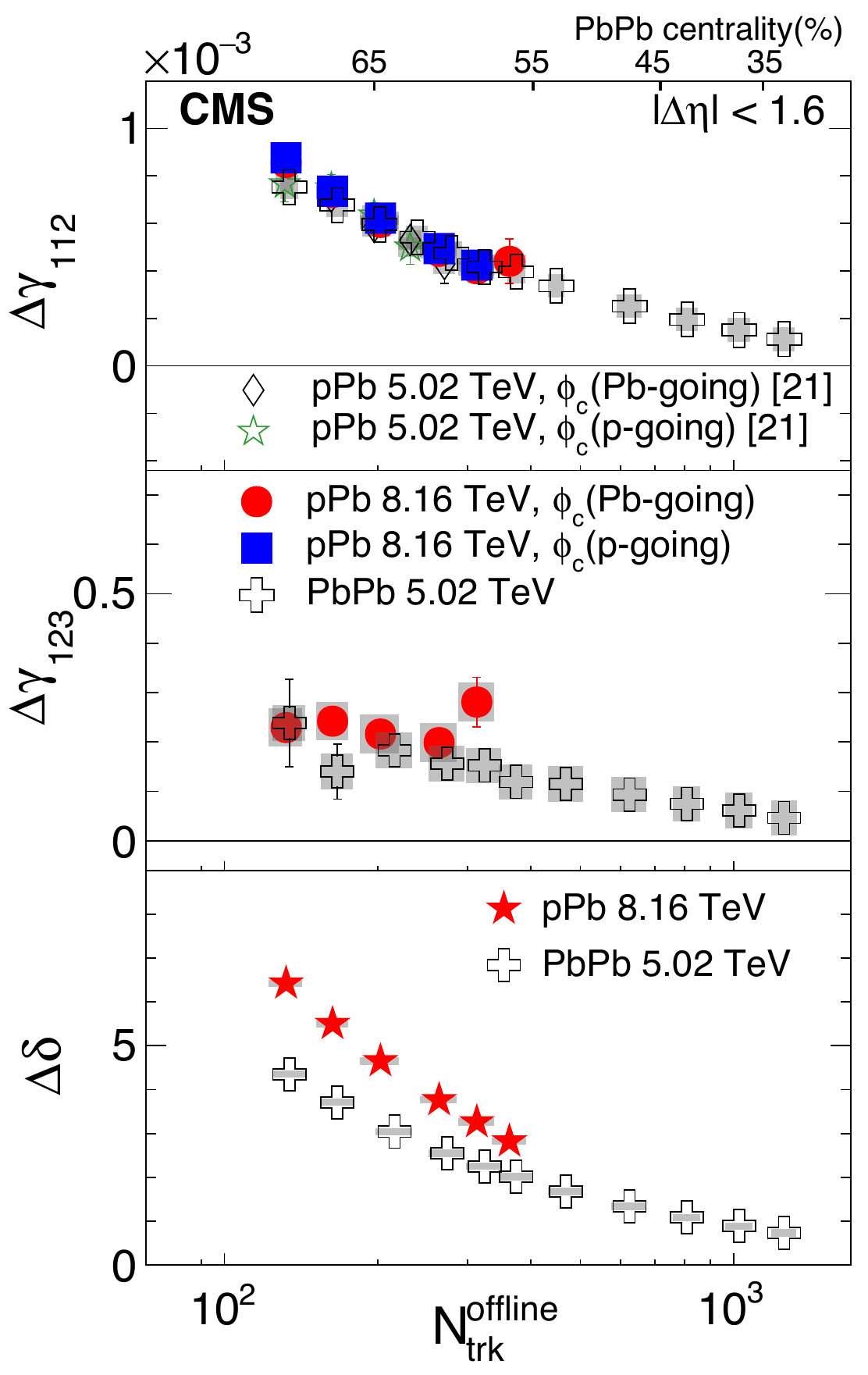} 
	\includegraphics[width=6.2cm]{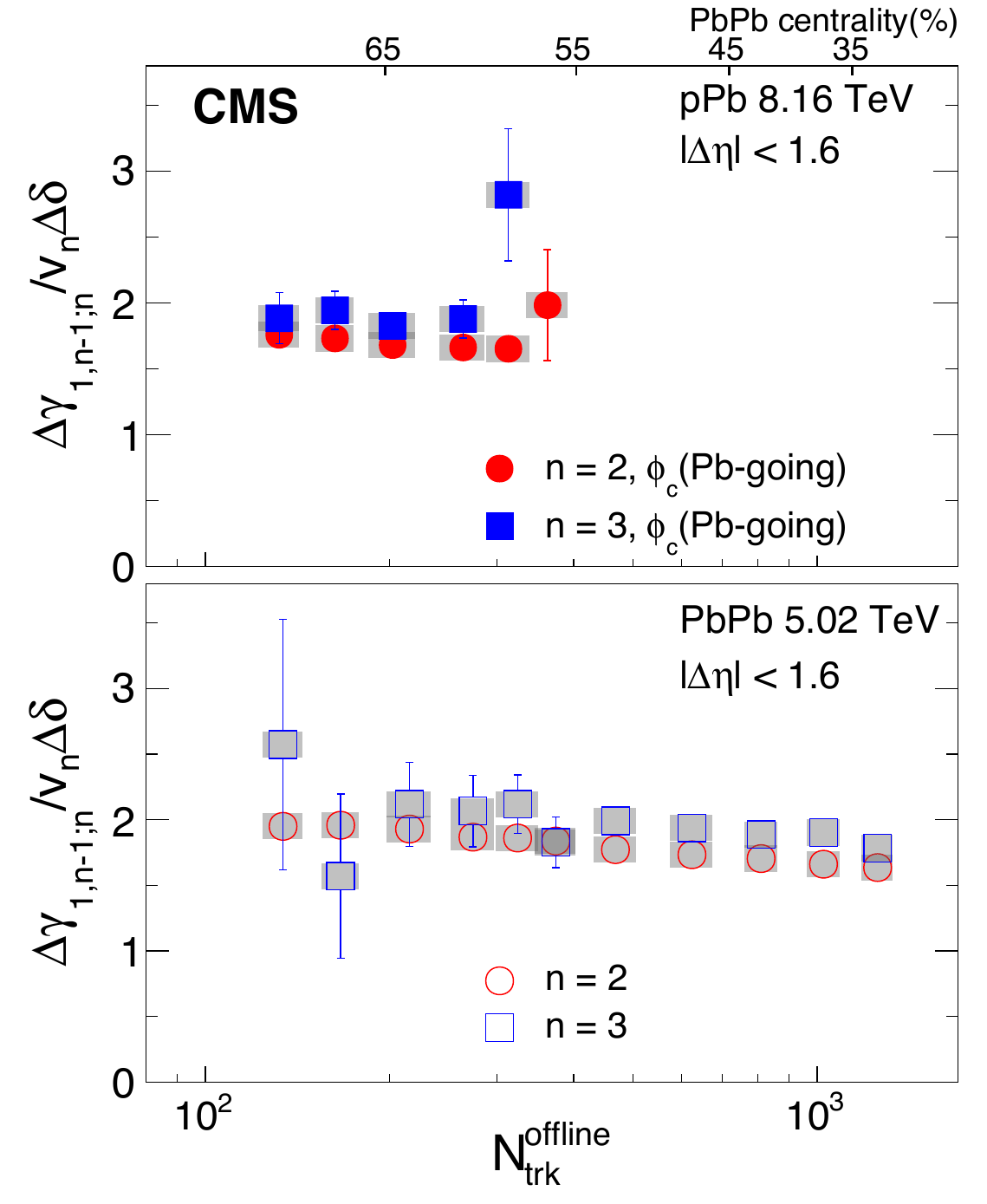} 
	\caption{(Color online)
		The $\dg_{112}$ (left top), $\dg_{123}$ (left middle), $\Delta\delta$ (left bottom), averaged over $|\Delta\eta|< 1.6$ as a function of
		$N^{\rm offline}_{\rm trk}$ in p+Pb and Pb+Pb collisions. The p+Pb results are obtained with particle $c$ from Pb- and p-going sides separately. 
		The ratio of $\dg_{112}$ and $\dg_{123}$ to the product of $v_{n}$ and $\Delta\delta$ in p+Pb collisions for the Pb-going direction (right top) 
		and Pb+Pb collisions (right bottom). 
		Statistical and systematic uncertainties are indicated by the error bars and shaded regions, respectively\cite{Sirunyan:2017quh}.
	}   
	\label{FG_CMSgm123}
\end{figure}

The CME-driven charge separation are expected along the magnetic field direction normal to the reaction plane, estimated by the second-order event plane ($\psi_{2}$).
The third-order event plane ($\psi_{3}$) are expected to be weakly correlated with $\psi_{2}$\cite{Aad:2014fla}, 
thus the CME-driven charge separation effect with respect to $\psi_{3}$ is expected to be negligible.
In light of $v_{2}$-dependent background-only scenario, where background can be expressed as Eq~\ref{EQ_ESE6}. 
A similar correlator ($\gamma_{123}$) with respect to third-order event plane ($\psi_{3}$) are constructed to study the background effects\cite{Sirunyan:2017quh}:
\begin{equation}
	\begin{split}
		\gamma_{123} & \equiv \mean{\cos(\phi_{\alpha}-2\phi_{\beta}-3\psi_{3})}  \\ 
		\gamma_{123}^{bkg} & =\kappa_{3}\mean{\cos(\phi_{\alpha}-\phi_{\beta})}\mean{\cos3(\phi_{\beta}-\psi_{3})}   \\
		    & =\kappa_{3} \delta v_{3}.  \\	
	\end{split}
	\label{EQ_ESE9}
\end{equation}
In the flow-dependent background-only scenario, the $\kappa_{2}$ and $\kappa_{3}$ mainly depend on particle kinematics and detector acceptance effects, 
and are expected to be similar, largely independent of harmonic event plane order. 
Fig.~\ref{FG_CMSgm123}(left) shows the $\dg$ ($\Delta\gamma_{112}$), $\Delta\gamma_{123}$, $\Delta\delta$ correlator as a 
function of multiplicity in p+Pb and Pb+Pb collisions. 
Fig.~\ref{FG_CMSgm123}(right) shows the ratio of the $\dg$ ($\Delta\gamma_{112}$) and $\Delta\gamma_{112}$ to the product of $v_{n}$ and $\Delta\delta$. 
The results show that the ratio is similar for $n$=2 and 3, and also similar between p+Pb and Pb+Pb collisions, indicating that the $\kappa_{3}$ is similar to $\kappa_{2}$. 
These results are consistent with the flow-dependent background-only scenario.

The event shape selection provides a very useful tool to study the background behavior of the $\dg$. 
All current experimental results at LHC suggest that the $\dg$ are strongly dependent on the $v_{2}$ and consistent with the flow-background only scenario. 
In summary, the $v_{2}$ independent contribution are estimated by different methods from STAR, ALICE and CMS, and current 
results indicate that a large contribution of the $\dg$ correlator is from the $v_{2}$ related background.

%%%%%%%%%%%%%%%%%%%%%%%%%%%%%%%%%%%%%%%%%%%%%%%%%%%%%%%%%%%%%%%%%%%%%%%%%%%%%%%%%%%%%%%%%%%%%%
\section{Isobaric collisions}
The CME is related to the magnetic field while the background is produced by $v_{2}$-induced correlations. 
In order to gauge differently the magnetic field relative to the $v_{2}$,
isobaric collisions and Uranium+Uranium collisions have been proposed\cite{Voloshin:2010ut}.
The isobaric collisions are proposed to study the two systems with similar $v_{2}$ but different magnetic field strength\cite{Voloshin:2010ut},
such as $\Ru$ and $\Zr$, which have the same mass number, but differ by charge (proton) number. 
Thus one would expected very similar $v_{2}$ at mid-rapidity in $\RuRu$ and $\ZrZr$ collisions, 
but the magnetic field, proportional to the nuclei electric charge, could vary by 10\%.
If the measured $\dg$ is dominated by the CME-driven charge separation, 
then the variation of the magnetic field strength between $\RuRu$ and $\ZrZr$ collision 
provides an ideal way to disentangle the signal of the chiral magnetic effect from $v_{2}$ related background, 
as the $v_{2}$ related backgrounds are  expected to be very similar between these two systems.

To test the idea of the isobaric collisions, Monte Carlo Glauber calculations\cite{Deng:2016knn,Huang:2017azw,Xu:2017zcn} of the 
spatial eccentricity ($\epsilon_{2}$) and the magnetic field strength\cite{Deng:2016knn,Xu:2017zcn} form $\RuRu$ and $\ZrZr$ collisions have been carried out. 
The Woods-Saxon spatial distribution is used\cite{Deng:2016knn}:
\begin{equation}
	\begin{split}
		\rho(r,\theta) = \frac{1}{1+\rm{exp}\{[r-R_{0}-\beta_{2}R_{0}Y_{2}^{0}(\theta)]/ \it{a} \}},
	\end{split}
	\label{EQ_ISO1}
\end{equation}
where $\rho^{0} = 0.16 fm^{-3}$ is the normal nuclear density, $R^{0}$ is the charge radius of the nucleus,
$a$ represent the surface diffuseness parameter.
$Y_{2}^{0}$ is the spherical harmonic. The parameter $a$ is almost identical for $\Ru$ and $\Zr$: $a\approx0.46$ fm.
$R_{0}=5.085$ fm and 5.020 fm are used for $\Ru$ and $\Zr$, and are used for both the proton and nucleon densities.
The deformity quadrupole parameter $\beta_{2}$ has large uncertainties; 
there are two contradicting sets of values from current knowledge\cite{Deng:2016knn}, 
$\beta_{2}(\Ru)=0.158$  and $\beta_{2}(\Zr)=0.080$ (referred to as case 1) vis a vis 
$\beta_{2}(\Ru)=0.053$ and $\beta_{2}(\Zr)=0.217$ (referred to as case 2).  
These would yield less than 2\% difference in $\epsilon_{2}$, hence a less than 2\% residual $v_{2}$ background, 
between $\RuRu$\ and $\ZrZr$\ collisions in the 20-60\% centrality range\cite{Deng:2016knn}.
In that centrality range, the mid-rapidity particle multiplicities are almost identical for $\RuRu$\ and $\ZrZr$\ collisions 
at the same energy\cite{Deng:2016knn,Xu:2017zcn}. 

\begin{figure}[htbp!]
	\centering 
	\includegraphics[width=6.2cm]{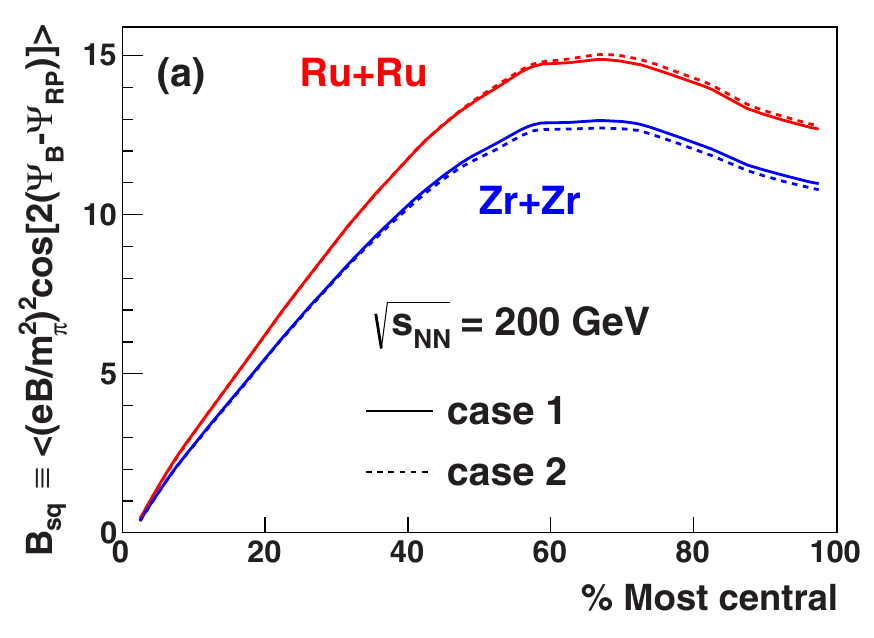} 
	\includegraphics[width=6.2cm]{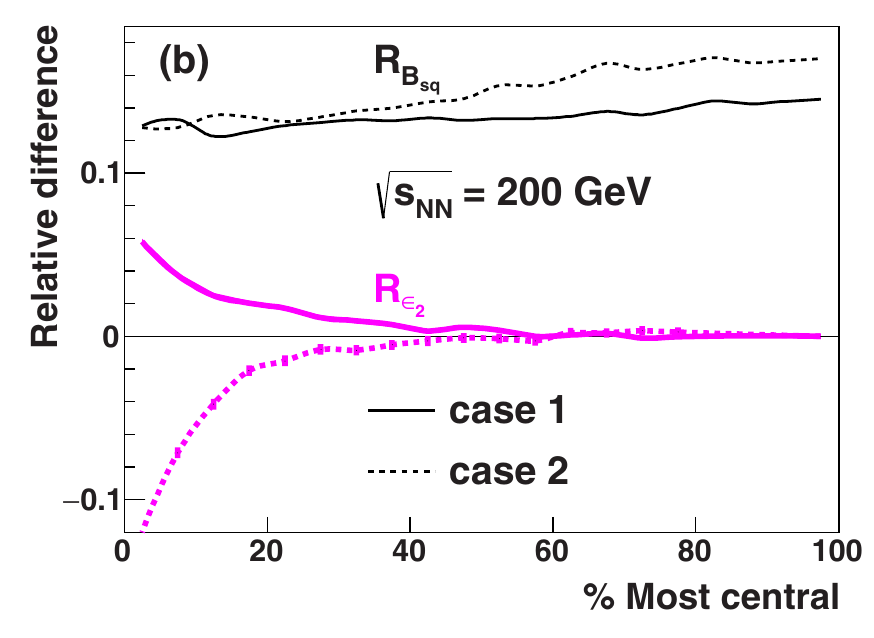} 
	\caption{(Color online)
		(a) Event-averaged initial magnetic field squared at the center of mass of the overlapping region, with correction from event-by-event
		fluctuation of magnetic field azimuthal orientation, for $\RuRu$ and $\ZrZr$ collisions at 200 GeV, and (b) their relative difference
		versus centrality. Also shown in (b) is the relative difference in the initial eccentricity. The solid (dashed) curves correspond to the parameter
		set of case 1 (case 2)\cite{Deng:2016knn}.
	}
	\label{FG_ISO1}
\end{figure}

The magnetic field strengths in $\RuRu$ and $\ZrZr$ collisions are calculated by using Lienard-Wiechert potentials 
alone with HIJING model taking into account the event-by-event azimuthal fluctuation of the magnetic field orientation\cite{Deng:2016knn,Deng:2012pc}.
HIJING model with the above two sets (case 1 and 2) of Woods-Saxon densities are simulated.
Fig.~\ref{FG_ISO1}(a) shows the calculation of the event-averaged initial magnetic field squared with correction from 
the event-by-event azimuthal fluctuation of the magnetic field orientation,  
\begin{equation}
	B_{sq} \equiv \mean{(eB/m_{\pi}^{2})^{2} \cos[2(\psi_{B} - \psi_{\rm{RP}})]}, 
\end{equation}
for the two	collision systems at 200 GeV. 
Fig.~\ref{FG_ISO1}(b) shows that the relative difference in $B_{sq}$, 
defined as $R_{B_{sq}}=2(B_{sq}^{Ru+Ru}-B_{sq}^{Zr+Zr})/(B_{sq}^{Ru+Ru}+B_{sq}^{Zr+Zr})$, between $\RuRu$ and $\ZrZr$ collisions is approaching
15\% (case 1) or 18\% (case 2) for peripheral events, and reduces to about 13\% (cases 1 and 2) for central events.
Fig.~\ref{FG_ISO1}(b) also shows the relative difference in the initial eccentricity 
($R_{\epsilon_{2}}=2(\epsilon_{2}^{Ru+Ru}-\epsilon_{2}^{Zr+Zr})/(\epsilon_{2}^{Ru+Ru}+\epsilon_{2}^{Zr+Zr})$), obtained from the Monte-Carlo Glauber
calculation. The relative difference in $\epsilon_{2}$ is practically zero for peripheral
events, and goes above (below) 0 for the parameter set of
case 1 (case 2) in central collisions. 
The relative difference in $v_{2}$ from $\RuRu$\ and $\ZrZr$\ collisions is expected to closely follow that in eccentricity, 
indicating the $v_{2}$-related backgrounds are almost the same (different within 2\%) for $\RuRu$ and $\ZrZr$ collisions in 20-60\% centrality range. 

\begin{figure}[htbp!]
	\centering 
	\includegraphics[width=5.8cm]{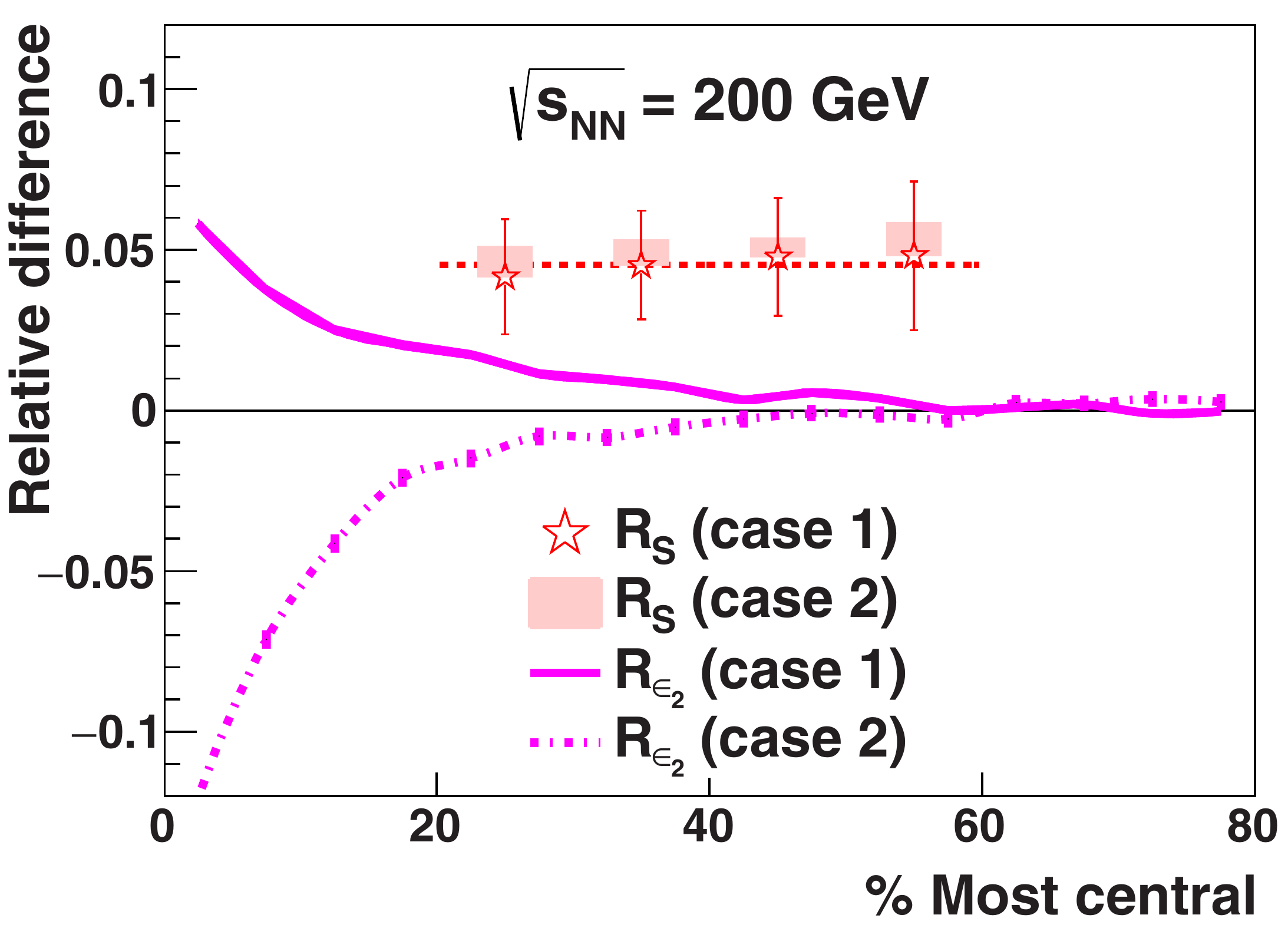} 
	\includegraphics[width=6.6cm]{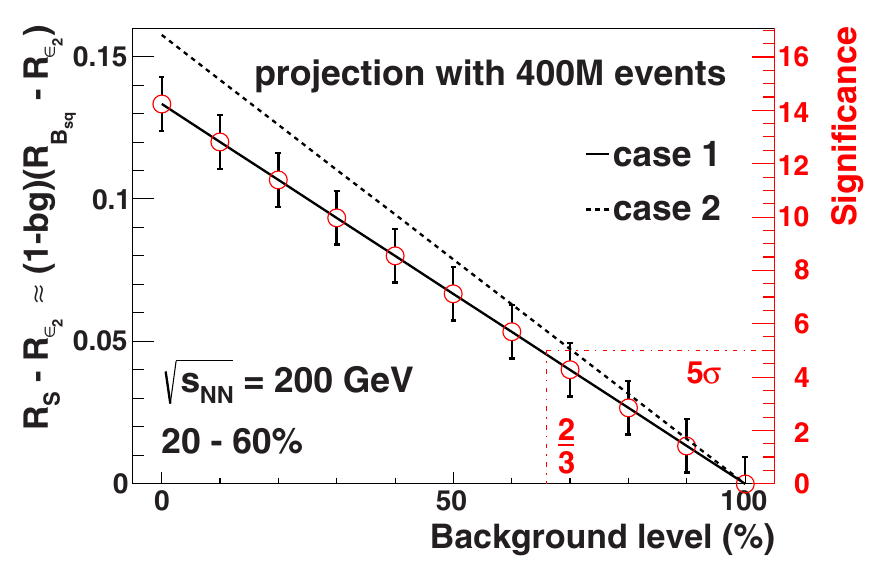} 
	\caption{(Color online)
		(Left) Estimated relative difference in $S = \dg \times N_{part}$ and in the initial eccentricity for $\RuRu$ and $\ZrZr$ collisions at 200 GeV. 
		(Right) Magnitude and significance of the relative difference in the CME signal between $\RuRu$ and $\ZrZr$,  
		$R_{S} - R_{\epsilon_{2}}$ as a function of the background level\cite{Deng:2016knn}.
	}
	\label{FG_ISO2}
\end{figure}

Based on the available experimental $\dg$ measurements in Au+Au collisions at 200 GeV and the calculated magnetic field 
strength and eccentricity difference between $\RuRu$ and $\ZrZr$ collisions, 
expected signals from the isobar collisions are estimated\cite{Deng:2016knn}: 
\begin{equation}
	\begin{split}
		&S = \dg \times N_{part}, \\	
		&S^{Ru+Ru} = \bar{S} \big[ (1-bg)\big(1+\frac{R_{B_{sq}}}{2}\big) +bg\big(1+\frac{R_{\epsilon_{2}}}{2} \big) \big], \\  
		&S^{Zr+Zr} = \bar{S} \big[ (1-bg)\big(1-\frac{R_{B_{sq}}}{2}\big) +bg\big(1-\frac{R_{\epsilon_{2}}}{2} \big) \big], \\
		&R_{S} = (1-bg)R_{B_{sq}} + bgR_{\epsilon_{2}}, \ bg \in [0,1], \\ 
	\end{split}
	\label{EQ_ISO2}
\end{equation}
where $S$ represents the $N_{part}$ scaled $\dg$ correlator ($N_{part}$ account for the dilution effect\cite{Abelev:2009ad,Ma:2011uma}).
The $bg$ is the $v_{2}$ related background fraction of the $\dg$ correlator. 
Fig.~\ref{FG_ISO2}(left) shows the $R_{S}$ with $400\times10^{6}$ events for each of the two collisions types, assuming $\frac{2}{3}$ of the $\dg$ comes from the $v_{2}$ related background, and compared with $R_{\epsilon_{2}}$.
Fig.~\ref{FG_ISO2}(right) shows the magnitude and significance of the projected relative difference 
between $\RuRu$ and $\ZrZr$ collisions as a function of the background level. 
With the given event statistics and assumed background level, the isobar collisions will give 5$\sigma$ significance.

\begin{figure}[htbp!]
	\centering 
	\includegraphics[width=5.8cm]{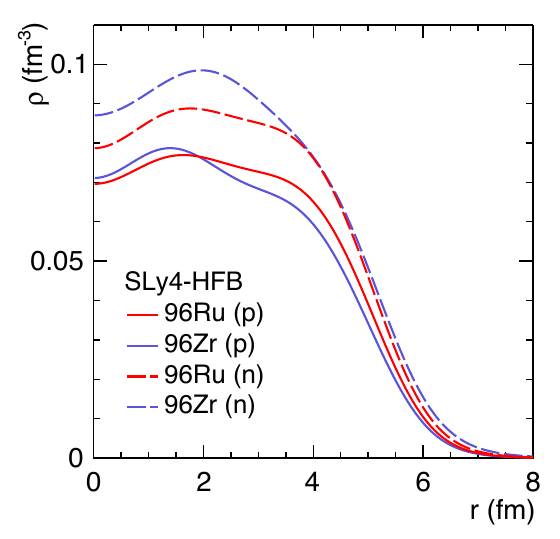} 
	\includegraphics[width=6.6cm]{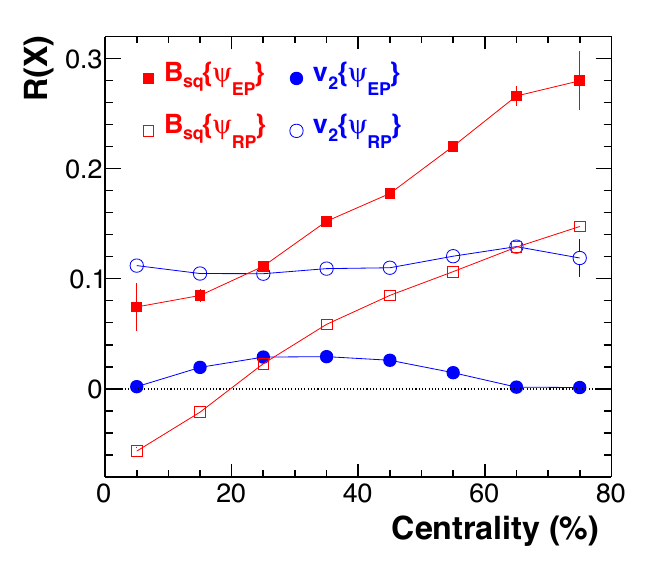} 
	\caption{(Color online)
		(Left) Proton and neutron density distributions of the $\Ru$ and $\Zr$ nuclei, assumed spherical, calculated by the EDF method\cite{Xu:2017zcn}. 
		(Right) Relative differences between $\RuRu$ and $\ZrZr$ collisions as functions of centrality in $v_{2}\{\psi\}$ and $Bsq\{\psi\}$ with respect to $\psiRP$ and $\psiEP$ from AMPT simulation with the densities from the left plot\cite{Xu:2017zcn}.
	}
	\label{FG_ISO3}
\end{figure}

The above estimates assume Woods-Saxon densities, identical for proton and neutron distributions.
Using the energy density functional (EDF) method with the well-known SLy4 mean field~\cite{Chabanat:1997un} including pairing correlations (Hartree-Fock-Bogoliubov, HFB approach)\cite{Wang:2016rqh,Bender:2003jk,ring1980nuclear}, 
assumed spherical, the ground-state density distributions for $\Ru$\ and $\Zr$ are calculated. The results are shown in Fig.~\ref{FG_ISO3}(left)\cite{Xu:2017zcn}. 
They show that protons in Zr are more concentrated in the core, while protons in Ru, 10\% more than in Zr, are pushed more toward outer regions. 
The neutrons in Zr, four more than in Ru, are more concentrated in the core but also more populated on the nuclear skin.
Fig.~\ref{FG_ISO3}(right) shows the relative differences between $\RuRu$ and $\ZrZr$ collisions as functions of centrality in $v_{2}\{\psi\}$ and $B_{sq}\{\psi\}$ with respect to $\psiRP$ and $\psiEP$ from AMPT simulation with the densities calculated by the EDF method.
Results suggest that with respect to $\psiEP$, the relative difference in $\epsilon_{2}$ and $v_{2}$ are as large as $\sim$3\%. 
With respect to $\psiRP$, the difference in $\epsilon_{2}$ and $v_{2}$ becomes even larger ($\sim$10\%), and the difference in $B_{sq}$ is only 0-15\%\cite{Xu:2017zcn}. 
These studies suggest that the premise of isobaric sollisions for the CME search may not be as good as originally anticipated, and 
could provide additional important guidance to the experimental isobaric collision program.

%%%%%%%%%%%%%%%%%%%%%%%%%%%%%%%%%%%%%%%%%%%%%%%%%%%%%%%%%%%%%%%%%%%%%%%%%%%%%%%%%%%%%%%%%%%%%%
\section{Uranium+Uranium collisions}
Isobaric collisions produce different magnetic field but similar $v_{2}$. One may produce on average different $v_{2}$ but same magnetic fields, this may be achieved by uranium+uranium collisions\cite{Voloshin:2010ut}. 
Unlike the nearly spherical nuclei of gold (Au), uranium (U) nuclei have a highly ellipsoidal shape. 
By colliding two uranium nuclei, there would be various collision geometries, such as the tip-tip or body-body collisions. 
In very central collisions, 
due to the particular ellipsoidal shape of the uranium nuclei, the overlap 
region would still be ellipsoidal in the body-body U+U collisions. 
This ellipsoidal shape of the overlap region would generate a finite elliptic flow, giving rise to the background in the $\dg$ measurements.
On the other hand,
the magnetic field are expected to vanish in the overlap region in those central body-body collisions. 
Thus in general the magnetic field driven CME signal will vanish in these very central collisions. 
By comparing central Au+Au collisions of different configurations, it may be possible to disentangle CME and background
correlations contributing to the experimental measured $\dg$ signal\cite{Voloshin:2010ut}. 
In 2012 RHIC ran U+U collisions.
Preliminary experimental results in central U+U have been compared with the results from central Au+Au\cite{Wang:2012qs,Tribedy:2017hwn}.
However, the geometry of the overlap region is much more complicated than initially anticipated, 
and the experimental systemic uncertainties are under further detailed investigation. 
So far there is no clear conclusion in term of the disentangle of the CME and $v_{2}$ related background from the preliminary experimental data yet.

%%%%%%%%%%%%%%%%%%%%%%%%%%%%%%%%%%%%%%%%%%%%%%%%%%%%%%%%%%%%%%%%%%%%%%%%%%%%%%%%%%%%%%%%%%%%%%
\section{Small system p+A or d+A collisions}
The small system p+A or d+A collisions provides a control experiment,
where the CME signal can be ``turned off'', but the $v_{2}$ related backgrounds still persist.
In non-central heavy-ion collisions, the $\psiPP$, although fluctuating\cite{Alver:2006wh}, is generally 
aligned with the reaction plane, 
thus generally perpendicular to $\Bvec$. 
The $\dg$ measurement is thus $\emph{entangled}$ by the two contributions: the possible CME and the $v_2$-induced background.
In small-system p+A or d+A collisions, however, the $\psiPP$ is determined purely by geometry fluctuations, 
uncorrelated to the impact parameter or the $\Bvec$ direction\cite{Khachatryan:2016got,Belmont:2016oqp,Tu:2017kfa}. 
As a result, any CME signal would average to zero in the $\dg$ measurements with respect to the $\psiPP$.
Background sources, on the other hand, contribute to small-system p+A or d+A collisions similarly as to heavy-ion collisions.
Comparing the small system p+A or d+A collisions to A + A collisions could thus further our understanding of the background issue in the $\dg$ measurements.

\begin{figure}[htbp!]
	\centering 
	\includegraphics[width=12.0cm]{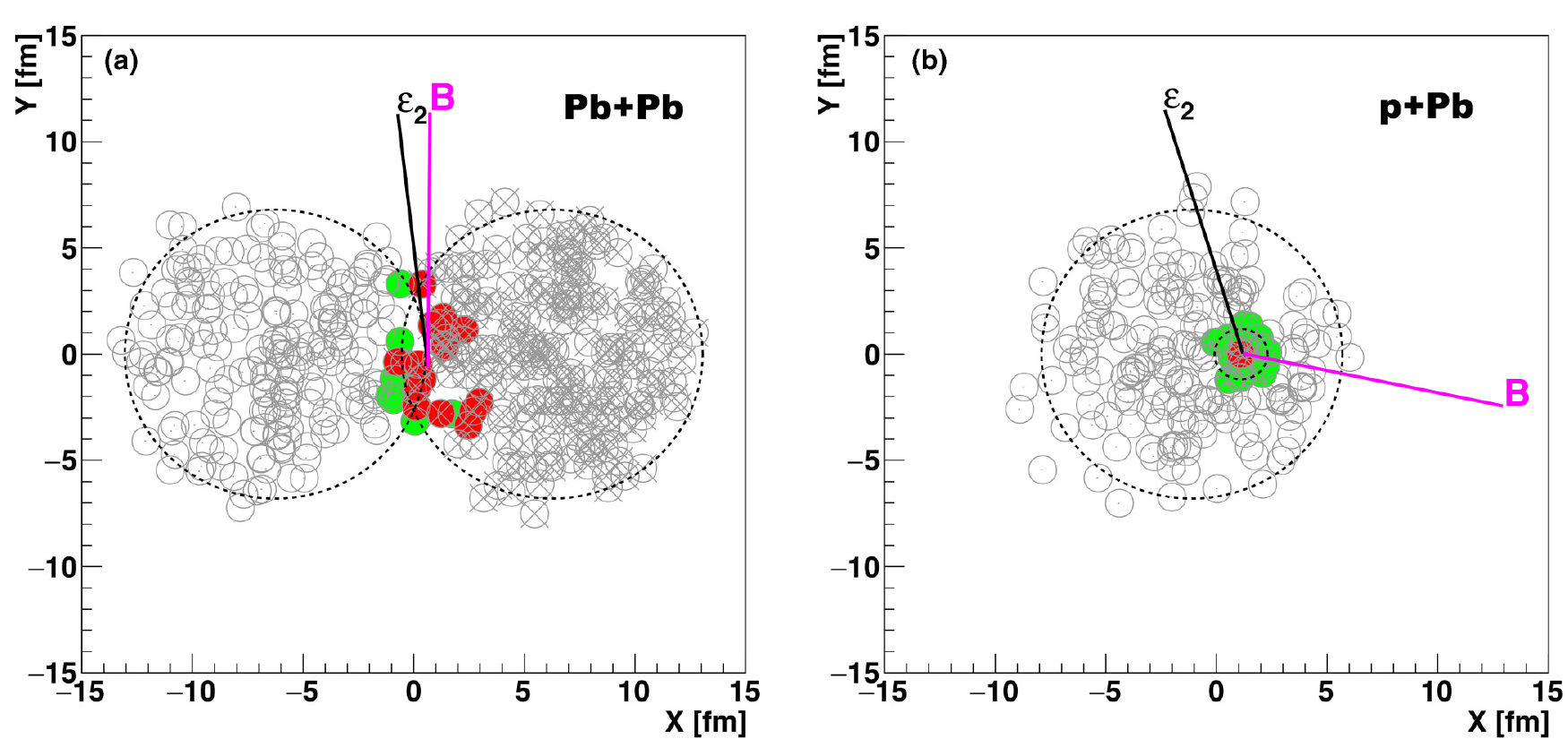} 
	\caption{(Color online)
		Single-event display from a Monte-Carlo Glauber event of a peripheral Pb+Pb (a) and a central p+Pb (b) collision at 5.02 TeV.
		The open gray [solid green (light gray)] circles indicate spectator nucleons (participating protons) traveling in the positive z direction, and the
		open gray [solid red (dark gray)] circles with crosses indicate spectator nucleons (participating protons) traveling in the negative z direction.
		In each panel, the calculated magnetic field vector is shown as a solid magenta line and the long axis of the participant eccentricity is shown as
		a solid black line\cite{Belmont:2016oqp}.
	}   
	\label{FG_SM1}
\end{figure}

Figure~\ref{FG_SM1} shows a single-event display from a Monte Carlo-Glauber event of a peripheral Pb+Pb (a) and a central p+Pb (b) collision at 5.02 TeV\cite{Belmont:2016oqp}.
In A+A collisions, due to the geometry of the overlap region, the eccentricity long axis are highly correlated with the impact parameter direction.
Meanwhile the magnetic field direction is mainly determined by the positions of the protons in the two colliding nucleus, 
which is also generally perpendicular to the impact parameter direction. 
Thus in A+A collisions, these two direction are highly correlated with each other. 
Consequently, the $\dg$ measurements are entangled with the $v_{2}$ background and possible CME signal.
While in p+A (Fig.~\ref{FG_SM1}, b) due to fluctuations in the positions of the nucleons, 
the eccentricity long axis and magnetic field direction are no longer correlated with each other. 
So the $\dg$ measurements in p+A collisions with respected to the eccentricity long axis (estimated by $\psiPP$) will lead to zero CME signal on average, 
and similarly for d+A collisions.

The recent $\dg$ measurements in small system p+Pb collisions from CMS have triggered a wave of 
discussions about the interpretation of the CME in heavy-ion collisions\cite{Khachatryan:2016got}.
The $\dg$ correlator signal from p+Pb is comparable to the signal from Pb+Pb collisions at similar multiplicities, 
which indicates significant background contributions in Pb+Pb collisions at LHC energy.

\begin{figure}[htbp!]
	\centering 
	\includegraphics[width=7.0cm]{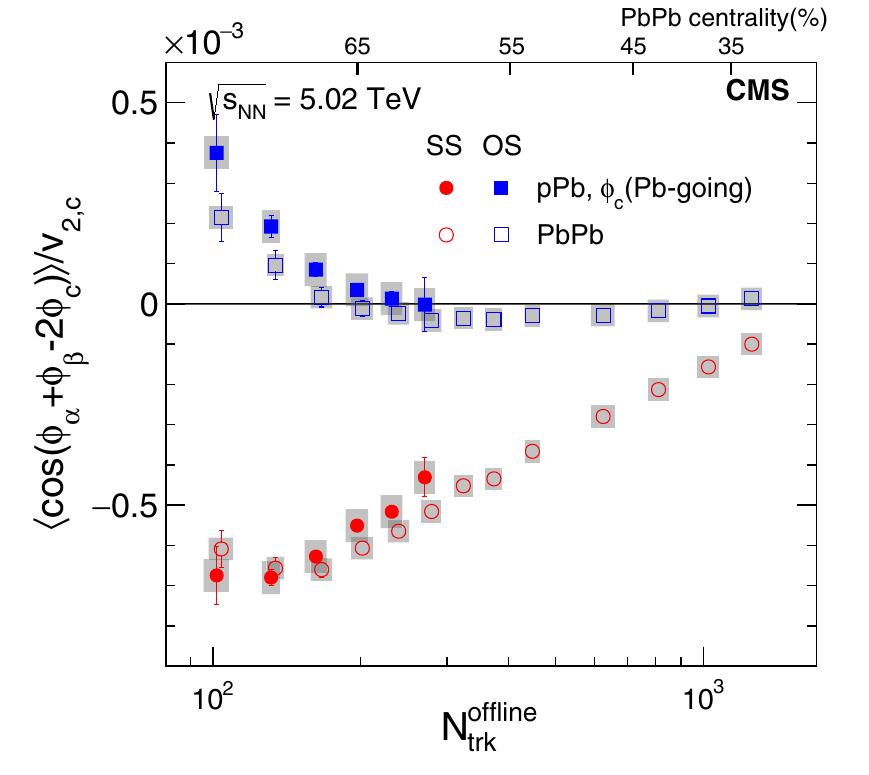} 
	\caption{(Color online)
		The opposite-sign and same-sign three-particle correlator
		averaged over $|\eta_{\alpha}-\eta_{\beta}| < 1.6$ as a function of $N^{\rm offline}_{\rm trk}$
		in p+Pb and Pb+Pb collisions at \sNN = 5.02 TeV from CMS collaboration.
		Statistical and systematic uncertainties are indicated by the error
		bars and shaded regions, respectively\cite{Khachatryan:2016got}.
	}   
	\label{FG_SM2}
\end{figure}

\begin{figure}[htbp!]
	\centering 
	\includegraphics[width=6.2cm]{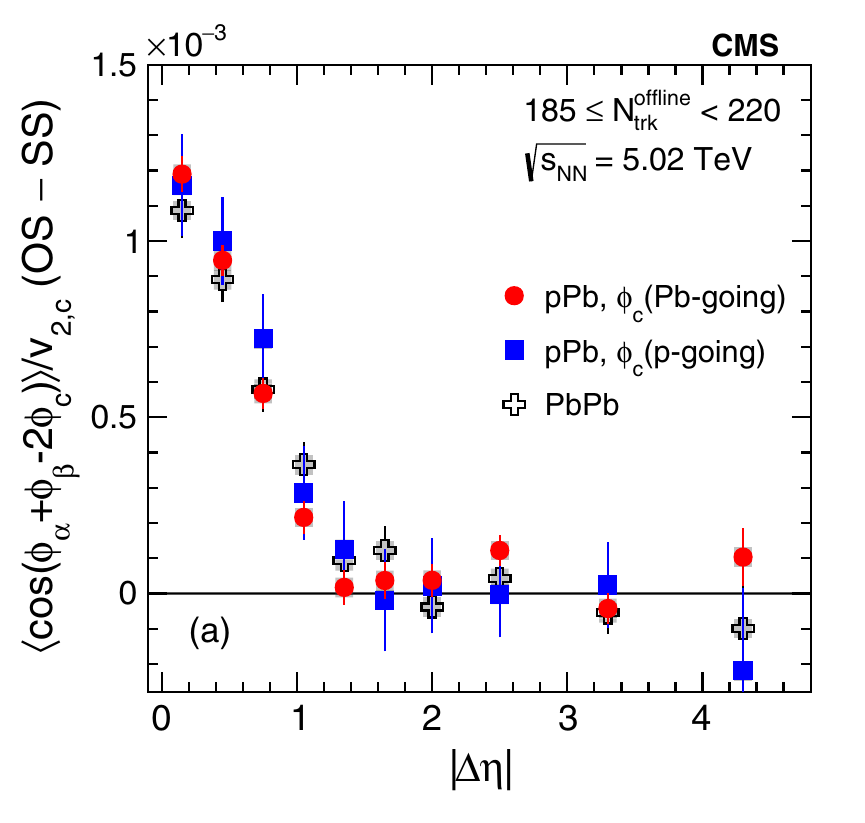} 
	\includegraphics[width=6.2cm]{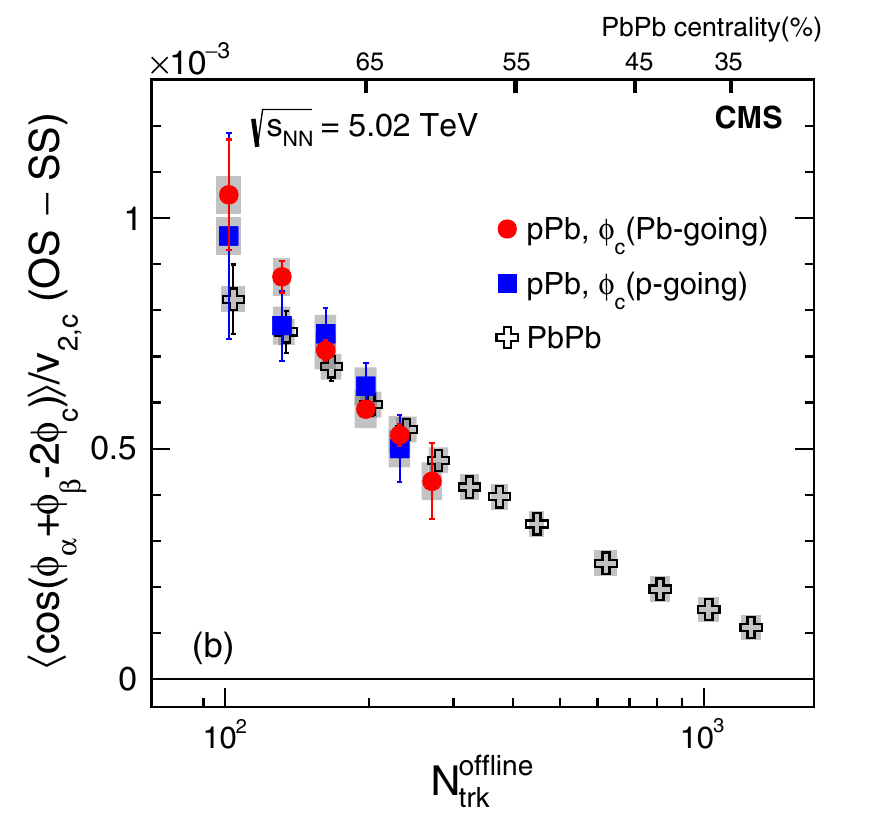} 
	\caption{(Color online)
		The difference of the opposite-sign and same-sign
		three-particle correlators (a) as a function of $|\eta_{\alpha}-\eta_{\beta}|$ for
		$185 \leq N^{\rm offline}_{\rm trk} < 220$ and (b) as a function of $N^{\rm offline}_{\rm trk}$, 
		averaged over $|\eta_{\alpha}-\eta_{\beta}| < 1.6$, in p+Pb and Pb+Pb collisions at
		\sNN = 5.02 TeV from CMS collaboration. The p-Pb results are obtained with particle $c$ from Pb- and p-going sides separately. 
		Statistical and systematic uncertainties are indicated by the error bars and shaded regions, respectively\cite{Khachatryan:2016got}.
	}   
	\label{FG_SM3}
\end{figure}

Figure~\ref{FG_SM2} shows the first $\dg$ measurements in small system p+A collisions from CMS, 
by using p+Pb collisions at 5.02 TeV compared with Pb+Pb at same energy. 
The results are plotted as a function of event charged-particle multiplicity ($N^{\rm offline}_{\rm trk}$).
The p+Pb and Pb+Pb results are measured in the same $N^{\rm offline}_{\rm trk}$ ranges up to 300.
The p+Pb results obtained with particle c in Pb-going forward direction. 
Within uncertainties, the SS and OS correlators in p+Pb and Pb+Pb collisions exhibit the
same magnitude and trend as a function of event multiplicity.
By taking the difference between SS and OS correlators, 
Fig~\ref{FG_SM3} shows the $|\Delta\eta|=|\eta_{\alpha}-\eta_{\beta}|$ and multiplicity dependence of $\dg$ correlator. 
The p+Pb and Pb+Pb data show similar $|\Delta\eta|$ dependence, decreasing with increasing $|\Delta\eta|$. 
The distributions show a traditional short range correlation structure, 
indicating the correlations may come from the hadonic stage of the collisions, 
while the CME is expected to be a long range correlation arising from the early stage.
The multiplicity dependence of $\dg$ correlator are also similar between p+Pb and Pb+Pb, decreasing as a function of $N^{\rm offline}_{\rm trk}$,
which could be understood as a dilution effect that falls with the inverse of event multiplicity\cite{Abelev:2009ad}.
There is a hint that slopes of the $N^{\rm offline}_{\rm trk}$ dependence in p+Pb and Pb+Pb are slightly different in Fig.~\ref{FG_SM3}(b),
which might be worth further investigation.
The similarity seen between high-multiplicity p+Pb and peripheral Pb+Pb collisions strongly suggests a common physical origin,
challenges the attribution of the observed charge-dependent correlations to the CME\cite{Khachatryan:2016got}. 

It is predicted that the CME would decrease with the collision energy due to the more rapidly decaying $B$ at
higher energies\cite{Kharzeev:2015znc,Deng:2012pc}. Hence, the similarity between small-system and heavy-ion collisions at the LHC may be expected, 
and the situation at RHIC could be different\cite{Kharzeev:2015znc}. 
Similar control experiments using p+Au and d+Au collisions are also performed at RHIC\cite{Zhao:2017wck,Zhao:2018pnk}.
\begin{figure}[htbp!]
	\centering 
	\includegraphics[width=6.2cm]{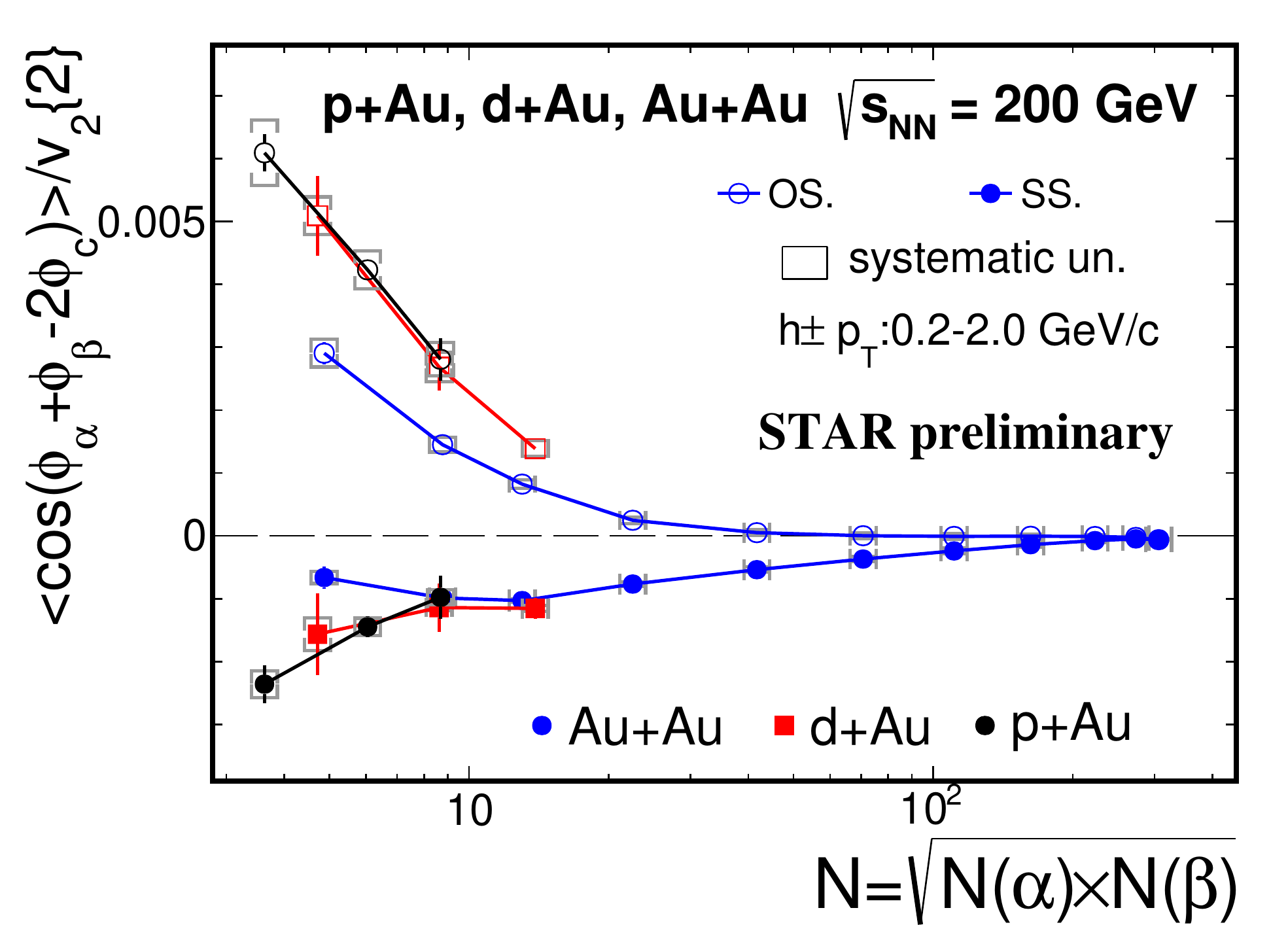} 
	\includegraphics[width=6.2cm]{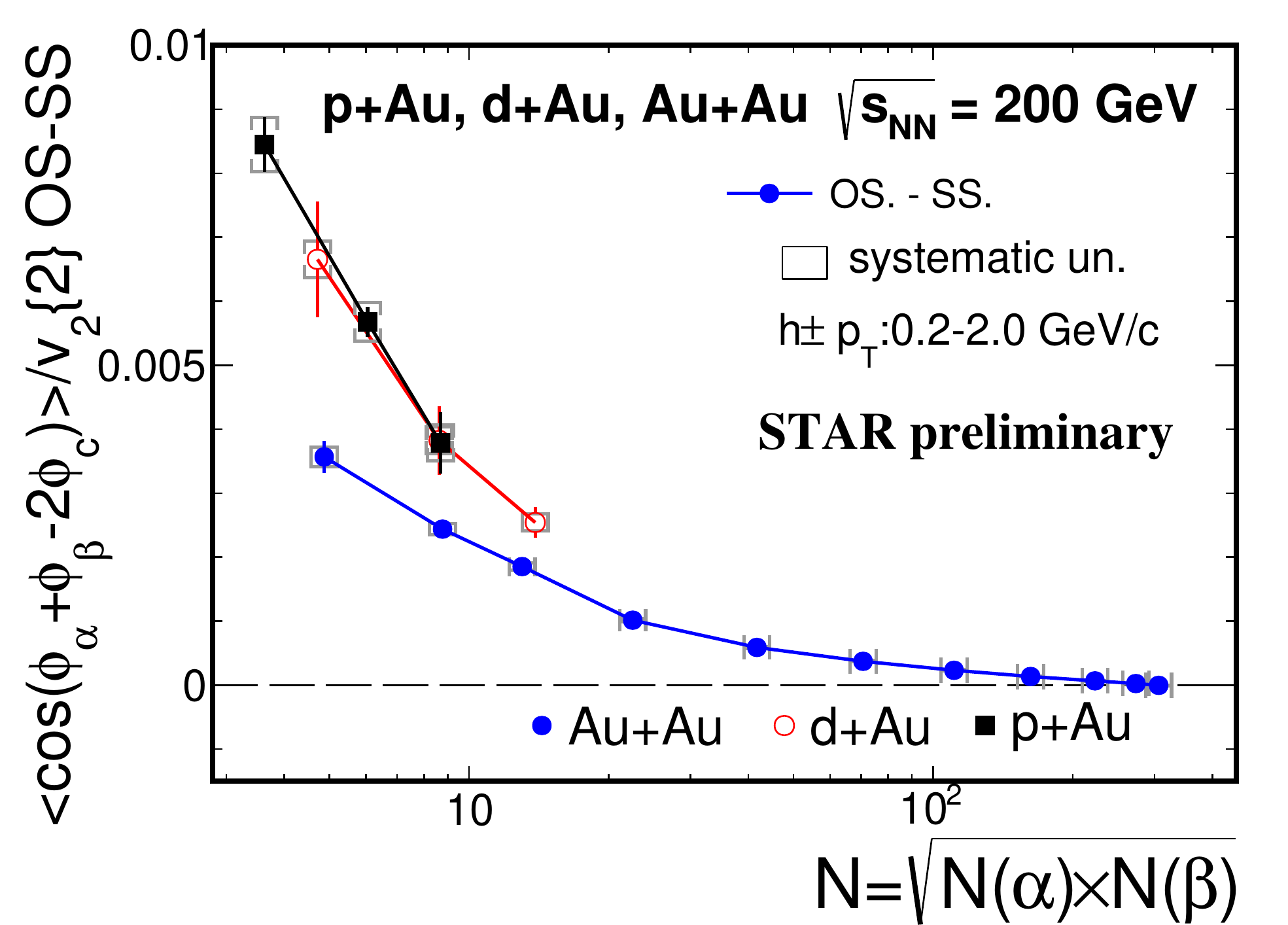} 
	\caption{(Color online)
		The preliminary $\gSS$, $\gOS$ (Left panel) and $\gdel$ (Right panel) correlators in p+Au and d+Au collisions as a function of multiplicity, 
		compared to those in Au+Au collisions at \sNN = 200 GeV from STAR collaboration. 
		Particles $\alpha$, $\beta$ and $c$	are from the TPC pseudorapidity coverage of $|\eta|<1$ with no $\eta$ gap applied.
		The $v_{2,c}\{2\}$ is obtained by two-particle cumulant with $\eta$ gap of $\Delta\eta > 1.0$. Statistical uncertainties
		are shown by the vertical bars and systematic uncertainties are shown by the caps\cite{Zhao:2017wck,Zhao:2018pnk}.
	}   
	\label{FG_SM4}
\end{figure}
Fig.~\ref{FG_SM4}(left) shows the $\gSS$ and $\gOS$ results as functions of particle multiplicity ($\mult$) in \pA\ and \dA\ collisions at $\snn=200$~GeV. 
Here $\mult$ is taken as the geometric mean of the multiplicities of particle $\alpha$ and $\beta$.
The corresponding \AuAu\ results are also shown for comparison. 
The trends of the correlator magnitudes are similar, decreasing with increasing $\mult$. 
The $\gSS$ results seem to follow a smooth trend in $\mult$ over all systems. 
The $\gOS$ results are less so; the small system data appear to differ somewhat from the heavy-ion data over the range in which they overlap in $\mult$. 
Similar to LHC, the small system $\dg$ results at RHIC are found to be comparable to Au+Au results at similar multiplicities (Fig.~\ref{FG_SM4}, right).
While in the overlapping $\mult$ range between \pdAu\ and \AuAu\ collisions, the $\dg$ data differ by $\sim$20-50\%.
This seems different from the LHC results where the \pPb\ and \PbPb\ data are found to be highly consistent with each other in the overlapping $\mult$ range\cite{Khachatryan:2016got}. 
However, the CMS \pPb\ data are from high multiplicity collisions, overlapping with \PbPb\ data in the 30-50\% centrality range, 
whereas the RHIC \pdAu\ data are from minimum bias collisions, overlapping with \AuAu\ data only in peripheral centrality bins. 
Since the decreasing rate of $\dg$ with $\mult$ is larger in \pdAu\ than in \AuAu\ collisions, 
the \pdAu\ data could be quantitatively consistent with the \AuAu\ data at large $\mult$ in the range of the 30-50\% centrality. 
It is interesting to note that this is similar to the observed difference in the slope of the $N^{\rm offline}_{\rm trk}$ dependence in p+Pb and Pb+Pb by CMS~\cite{Khachatryan:2016got} as mentioned previously. 
Considering these observations, the similarities in the RHIC and LHC data regarding the comparisons between small-system and heavy-ion collisions are astonishing.

\begin{figure}[htbp!]
	\centering 
	\includegraphics[width=7.0cm]{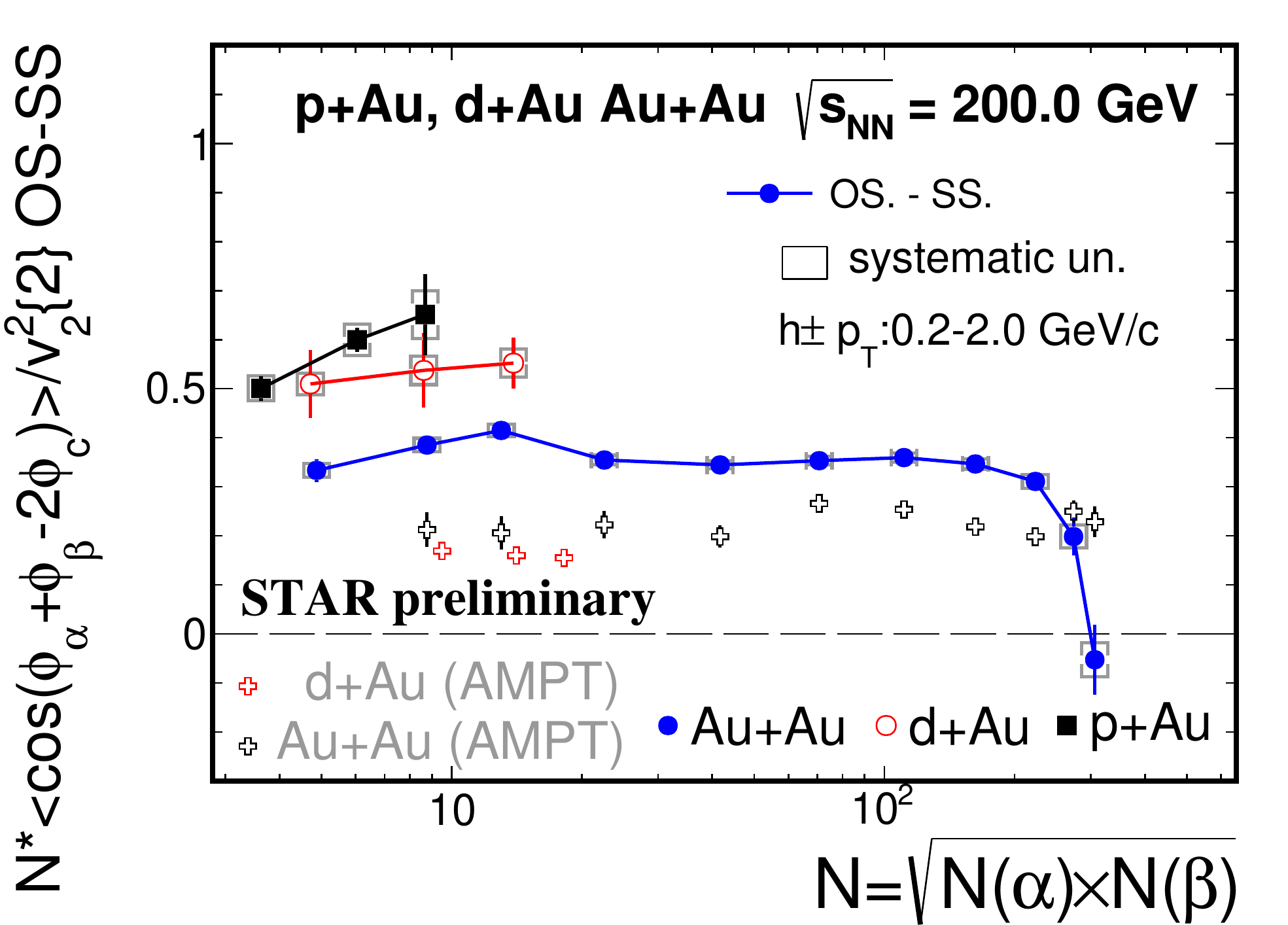} 
	\caption{(Color online)
		The scaled three-particle correlator difference in p+Au and d+Au collisions as a function of $\mult$, 
		compared to those in Au+Au collisions at \sNN = 200 GeV from the preliminary STAR data. 
		AMPT simulation results for d+Au and Au+Au are also plotted for comparison\cite{Zhao:2017wck,Zhao:2018pnk}.
	}   
	\label{FG_SM5}
\end{figure}

Since the \pA\ and \dA\ data are all backgrounds, the $\dg$ should be approximately proportional to the averaged $v_2$ of the background sources, 
and in turn, the $v_2$ of final-state particles. 
It should also be proportional to the number of background sources, 
and, because $\dg$ is a pair-wise average, inversely proportional to the total number of pairs as the dilution effect. 
The number of background sources likely scales with multiplicity, so the $\dg\propto v_2/\mult$.
Therefore, to gain more insight, the $\dg$ was scaled by $\mult/v_2$:
\begin{equation}
	\dgscale=\dg\times\mult/v_2\,.
	\label{eq:scale}
\end{equation}
Fig.~\ref{FG_SM5} shows the scaled $\dgscale$ as a function of $\mult$ in \pA\ and \dA\ collisions, and compares that to in \AuAu\ collisions. 
AMPT simulation results for d+Au and Au+Au are also plotted for comparison. The AMPT simulations can account for about $2/3$ of the STAR data, 
and are approximately constant over $\mult$. 
The $\dgscale$ in \pA\ and \dA\ collisions are compatible or even larger than that in \AuAu\ collisions. 
Since in \pA\ and \dA\ collisions only the background is present, the data suggest that the peripheral \AuAu\ measurement may be largely, if not entirely, background.
For both small-system and heavy-ion collisions, the $\dgscale$ is approximately constant over $\mult$. 
It may not be strictly constant because
the correlations caused by decays ($\dg_{\rm bkgd}\propto\mean{\cos(\alpha+\beta-2\phires)}\times\vres$), 
depends on the $\mean{\cos(\alpha+\beta-2\phires)}$ which is determined by the parent kinematics and can be somewhat $\mult$-dependent. 
Given that the background is large, suggested by the \pA\ and \dA\ data, the approximate $\mult$-independent $\dgscale$ in \AuAu\ collisions is consistent with the background scenario.

Due to the decorrelation of the $\psiPP$ and the magnetic field direction in small system \pdA\ collisions, 
the comparable $\dg$ measurements (with respect to the $\psiPP$) in small system \pdA\ collisions and in A+A collisions at the same energy from LHC/RHIC suggests 
that there is significant background contribution in the $\dg$ measurements in A+A collisions, 
where the $\dg$ measurements (with respect to the $\psiPP$) in small system \pdA\ collisions are all backgrounds.    
While, by considering the fluctuating proton size, Monte Carlo Glauber model calculation shows that there could be significant correlation between the magnetic field
direction and $\psiPP$ direction in high multiplicity p+A collisions, even though the magnitude of the correlation is still much smaller than in
A+A collisions. Those calculations may indicate possibilities of studying the chiral magnetic effect in small systems\cite{Kharzeev:2017uym,Zhao:2017rpf}.

The decorrelation of the $\psiPP$ and the magnetic field direction in small system \pdA\ collisions provides not only a way to ``turn off'' the CME signal, 
but also a way to ``turn off'' the $v_{2}$-related background.
The background contribution to the $\dg$ measurement with respect to the magnetic field direction would 
average to zero due to this decorrelation effect in system \pdA\ collisions.
So the key question is weather we could measure a direction that possibly accesses the magnetic field direction. 
The magnetic field is mainly generated by spectator protons and 
therefore experimentally best measured by the 1st-order harmonic plane ($\psi_{1}$) using the spectator neutrons.
\begin{figure}[htbp!]
	\centering 
	\includegraphics[width=7.0cm]{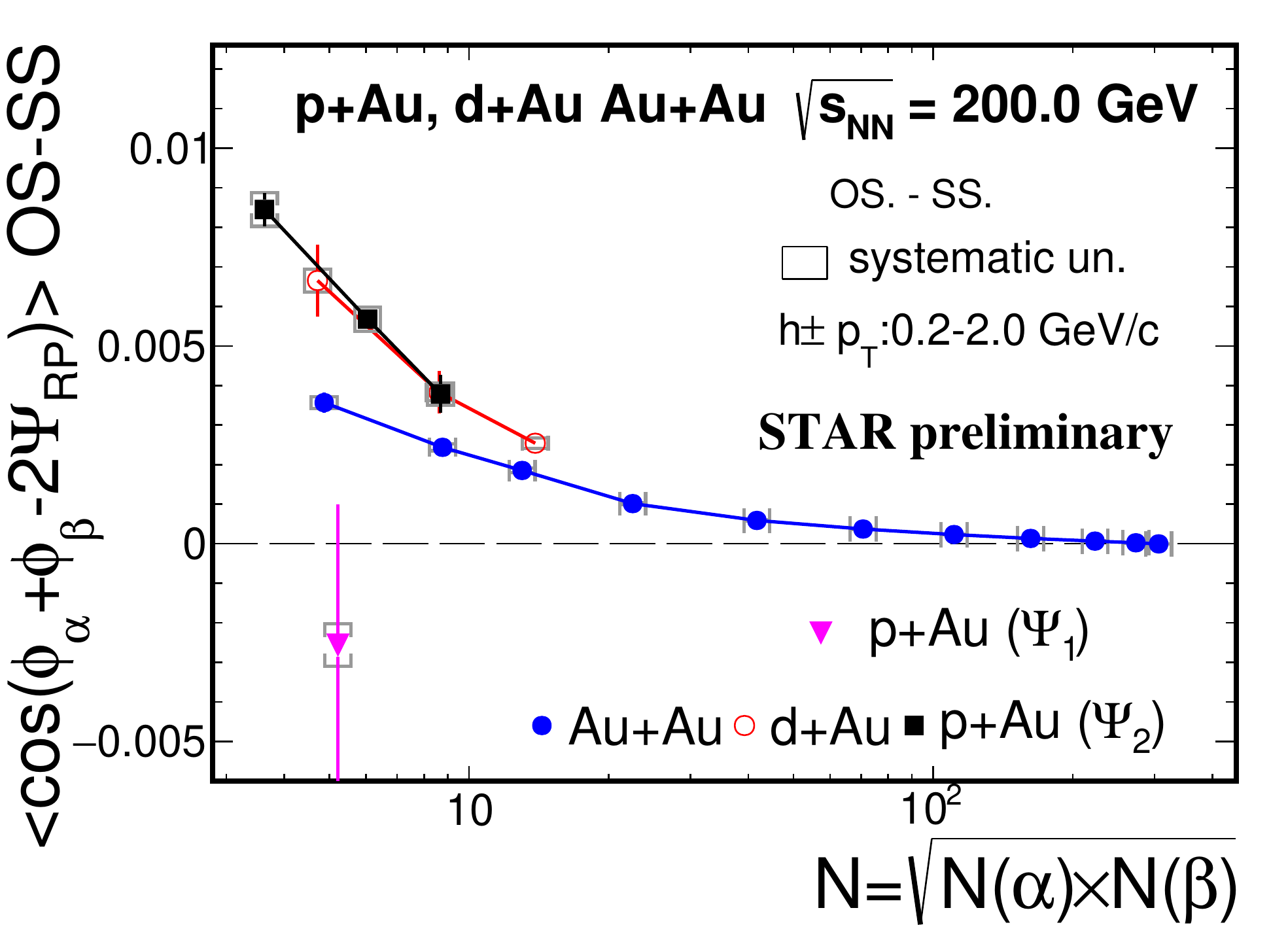} 
	\caption{(Color online)
		The preliminary $\dg$ correlator in p+Au collisions with respect to $\psi_{1}$ of spectator neutrons measured by the ZDC-SMD,
		compared to the $\dg$ measured with respected to $\psi_{2}$ in \pdAu\ and Au+Au collisions at \sNN = 200 GeV from STAR\cite{Zhao:2017p1}. 
	}   
	\label{FG_SM6}
\end{figure}
Fig.~\ref{FG_SM6} shows the preliminary $\dg$ measurement in p+Au collisions with respect to $\psi_{1}$ of spectator neutrons measured by the
shower maximum detectors of zero-degree calorimeters (ZDC-SMD) from STAR. The measurement is currently consistent with zero with large uncertainty\cite{Zhao:2017p1}.
In the future with improved experimental precision, this could possibly provide an excellent way to search for CME in small systems.

%%%%%%%%%%%%%%%%%%%%%%%%%%%%%%%%%%%%%%%%%%%%%%%%%%%%%%%%%%%%%%%%%%%%%%%%%%%%%%%%%%%%%%%%%%%%%%
\section{Measurement with respect to reaction plane}
\begin{figure}[htbp!]
	\centering 
	\includegraphics[width=7.0cm]{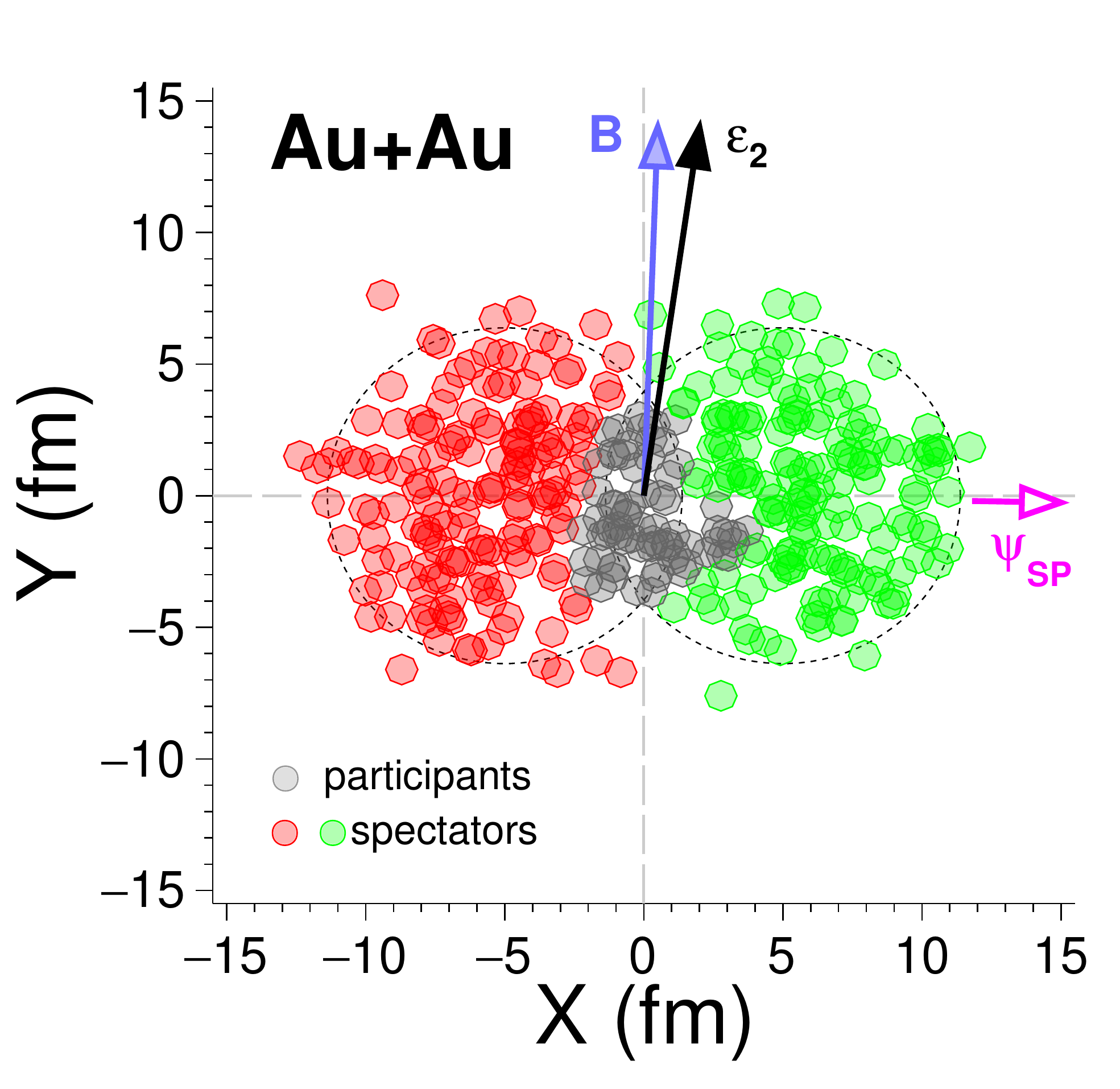} 
	\caption{(Color online)
		Single-event display from a Monte-Carlo Glauber event of a mid-central Au+Au collision at 200 GeV.
		The gray markers indicate participating nucleons, and the red (green) markers indicate the spectator nucleons traveling in positive (negative) z direction 
		The blue arrow indicates the magnetic field direction. The long axis of the participant eccentricity is shown as the black arrow.
		The magenta arrow shows the direction determined by spectator nucleons.
	}
	\label{FG_RP1}
\end{figure}

Again, one important point is that the CME-driven charge separation is along the magnetic field direction ($\psi_{B}$), 
different from the participant plane ($\psiPP$).
The major background to the CME is related to the elliptic flow anisotropy ($v_{2}$), determined by the participant geometry, 
therefore the largest with respect to the $\psiPP$.
The $\psi_{B}$ and $\psiPP$ in general correlate with the $\psiRP$, the impact parameter direction, therefore correlate to each other. 
While the magnetic field is mainly produced by spectator protons, their positions fluctuate, thus $\psi_{B}$ is not always perpendicular to the $\psiRP$. 
The position fluctuations of participant nucleons and spectator protons are independent, thus $\psiPP$ and $\psi_{B}$ fluctuate independently about $\psiRP$. 
Fig.~\ref{FG_RP1} depicts the display from a single Monte-Carlo Glauber event in mid-central Au+Au collision at 200 GeV.

The eccentricity of the transverse overlap geometry is by definition $\epsilon_{2}\{\psiPP\} \equiv \mean{\epsilon_{2}\{\psiPP\}_{evt}}$. 
The overlap geometry averaged over many events is an ellipse with its short axis being along the $\psiRP$; 
its eccentricity is $\mean{\epsilon_{2}\{\psiPP\}_{evt} \cos2(\psiPP - \psiRP)}$ and 
\begin{equation}
	\begin{split}
		& a^{PP}(\epsilon_{2}) \equiv \epsilon_{2}\{\psiRP\}/\epsilon_{2}\{\psiPP\} \approx a^{PP},   \\
		& a^{PP} \equiv \mean{\cos2(\psiPP - \psiRP)}. 
	\end{split}
	\label{EQ_RP1}
\end{equation}
The magnetic field strength with respect to a direction $\psi$ is: $B_{sq}\{\psi\} \equiv \mean{(eB/m_{\pi}^{2})^{2}\cos2(\psi_{B}-\psi)}$. And  
\begin{equation}
	\begin{split}
		a^{PP}_{B_{sq}} \equiv B_{sq}\{\psiPP\}/B_{sq}\{\psiRP\} \approx a^{PP}. 
	\end{split}
	\label{EQ_RP2}
\end{equation}

The relative difference of the eccentricity ($\epsilon_{2}$) or magnetic field strength ($B_{sq}$)  with respect to $\psiPP$ and $\psiRP$ are defined below:

\begin{equation}
	\begin{split}
		R^{PP}(X)            & \equiv 2\cdot \frac{X\{\psiRP\}-X\{\psiPP\}}{X\{\psiRP\}-X\{\psiPP\}}, \  X = \epsilon_{2}, or B_{sq},  \\
		R^{PP}(\epsilon_{2}) & \equiv -2(1-a^{PP}_{\epsilon_{2}})/(1+a^{PP}_{\epsilon_{2}}) \approx -R_{PP},  \\
		R^{PP}(B_{sq})       & \equiv  2(1-a^{PP}_{B_{sq}})/(1+a^{PP}_{B_{sq}}) \approx R_{PP},
	\end{split}
	\label{EQ_RP3}
\end{equation}
where
\begin{equation}
	\begin{split}
	    R^{PP} \equiv  2(1-a^{PP})/(1+a^{PP}). 
	\end{split}
	\label{EQ_RP4}
\end{equation}

The $\psiPP$ and $\epsilon_{2}$ are not experimentally measured. Usually the event plane ($\psiEP$) reconstructed from final-state particles is used as a proxy for $\psiPP$. 
$v_{2}$ can be used as a proxy for $\epsilon_{2}$: 
\begin{equation}
	\begin{split}
		a^{EP}_{v_{2}} \equiv v_{2}\{\psiRP\}/v_{2}\{\psiEP\} \approx a^{EP}. 
	\end{split}
	\label{EQ_RP5}
\end{equation}
Although a theoretical concept, the $\psiRP$ may be assessed by Zero-Degree Calorimeters (ZDC) measuring spectator neutrons\cite{Reisdorf:1997flow,Abelev:2013cva,Adamczyk:2016eux}. 
Similar to Eq.\ref{EQ_RP1},\ref{EQ_RP2},\ref{EQ_RP3},\ref{EQ_RP4}, these relations hold by replacing the $\psiPP$ with $\psiEP$. For example,
\begin{equation}
	\begin{split}
	    R^{EP} \equiv  2(1-a^{EP})/(1+a^{EP}). 
	\end{split}
	\label{EQ_RP6}
\end{equation}

\begin{figure}[htbp!]
	\centering 
	\includegraphics[width=12.5cm]{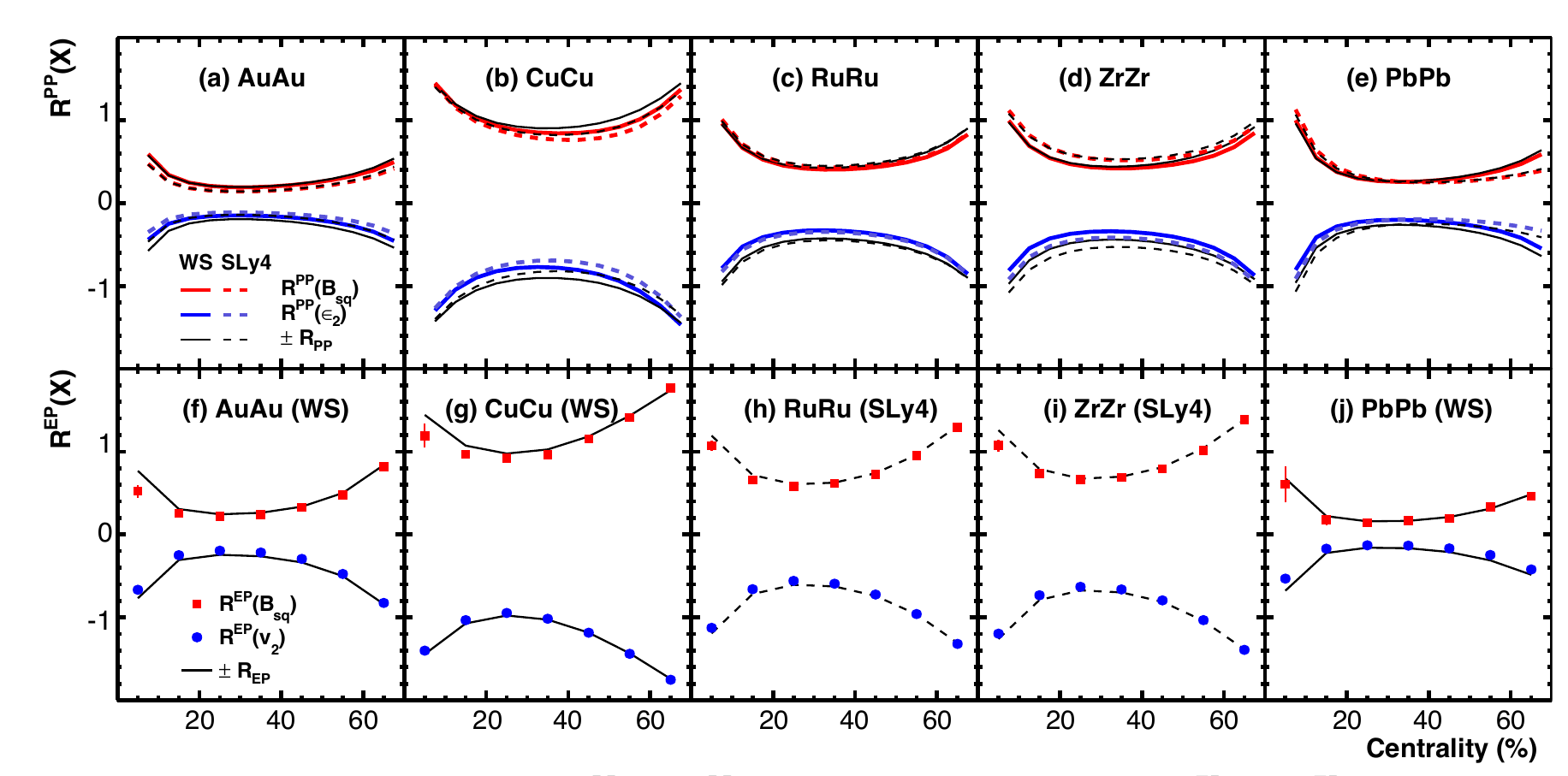} 
	\caption{(Color online)
		Relative differences $R^{PP}(\epsilon_{2})$, $R^{PP}(B_{sq})$ from Monte Carlo Glauber model (upper panel) and 
		$R^{EP}(v_{2})$, $R^{EP}(B_{sq})$ from AMPT (lower panel) for (a,f) Au+Au, (b,g) Cu+Cu, (c,h) Ru+Ru, and 
		(d,i) Zr+Zr at RHIC, and (e,j) Pb+Pb at the LHC. 
		Both the Woods-Saxon and EDF-calculated\cite{Xu:2017zcn} densities are shown for the Monte Carlo Glauber calculations, 
		while the used density profiles are noted for the AMPT results\cite{Xu:2017qfs}.
	}
	\label{FG_RP2}
\end{figure}

Figure~\ref{FG_RP2}(upper panel) shows $R^{PP}(\epsilon_{2})$ and $R^{PP}(B_{sq})$ calculated by a Monte Carlo Glauber model\cite{Xu:2014ada,Zhu:2016puf} 
for Au+Au, Cu+Cu, Ru+Ru, Zr+Zr collisions at RHIC and Pb+Pb collisions at the LHC. 
The results are compared to the corresponding $\pm R^{PP}$. 
These numbers agree with each other, indicating good approximations used in Eq.~\ref{EQ_RP1},\ref{EQ_RP2}.  
Fig.~\ref{FG_RP2}(lower panel) shows $R^{EP}(v_{2})$ and $R^{EP}(B_{sq})$ calculated from AMPT simulation\cite{Lin:2004en,Lin:2001zk}. 
Again, good agreements are found between $R^{EP}(v_{2})$, $R^{EP}(B_{sq})$ and $\pm R^{EP}$.
Both show the opposite behavior of $R^{PP(EP)}(\epsilon_{2}(v_{2}))$ and $R^{PP(EP)}(B_{sq})$, which approximately equal to $\pm R_{PP(EP)}$.

\begin{figure}[htbp!]
	\centering 
	\includegraphics[width=6.7cm]{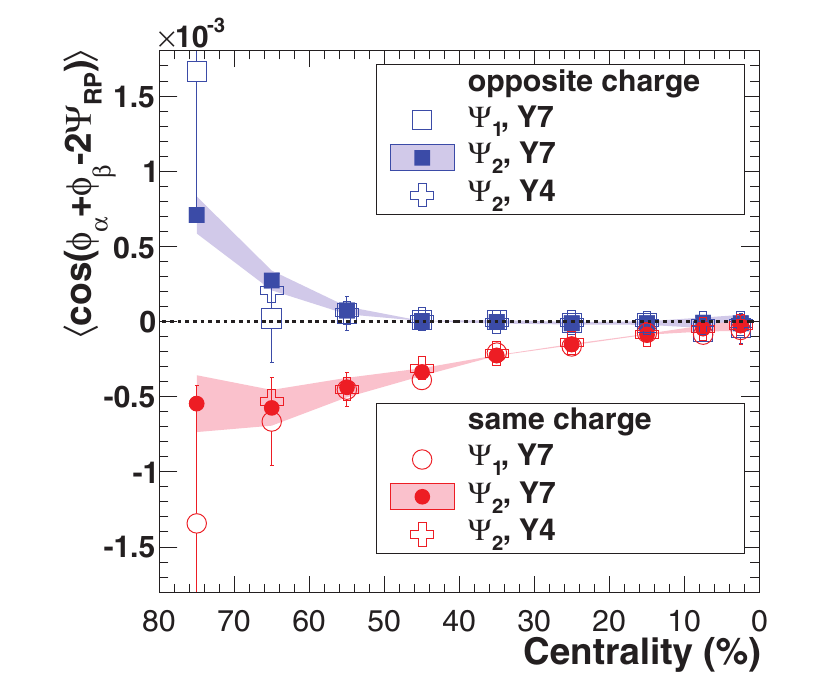} 
	\includegraphics[width=5.7cm]{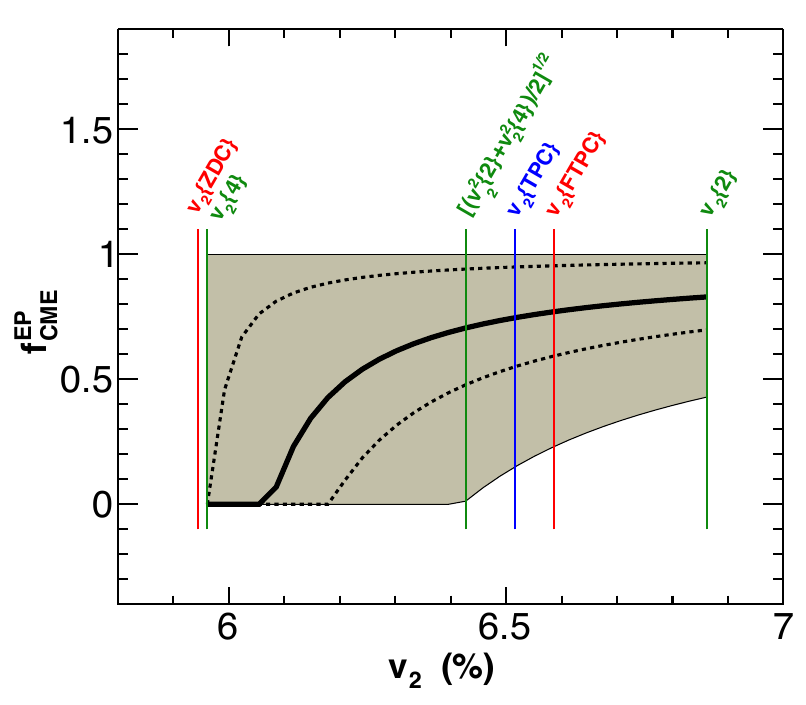} 
	\caption{(Color online)
		(Left) $\gamma$ measured with $\psi_{1}$ and $\psi_{2}$ vs. centrality in Au+Au collisions at \sNN = 
		200 GeV from STAR\cite{Adamczyk:2013hsi,Abelev:2009ac,Abelev:2009ad}. 
		The Y4 and Y7 represent the results from the 2004 and 2007 RHIC run. 
		Shaded areas for the results measured with $\psi_{2}$ represent the systematic uncertainty of the event plane determination. 
		Systematic uncertainties for the results with respect to $\psi_{1}$ are negligible compared to the statistical ones shown.
		(Right) The extracted fraction of CME contribution\cite{Xu:2017qfs} in the $\dg\{\psi_{2}\}$\cite{Abelev:2009ac,Abelev:2009ad,Adamczyk:2013hsi} 
		measurement in the 20-60\% centrality Au+Au collisions vs. ``true'' $v_{2}$;
		the gray area indicates the $\pm1\sigma$ statistical uncertainty, dominated by that in $\dg\{\psi_{1}\}$\cite{Adamczyk:2013hsi}. 
		The dashed curves would be the new $\pm1\sigma$ uncertainty with ten-fold increase in statistics.
	}
	\label{FG_RP3}
\end{figure}

The $\dg$ variable contains CME signal and the $v_{2}$-induced background:
\begin{equation}
	\begin{split}
		\dg\{\psi\} = \rm{CME}(B_{sq}\{\psi\}) + \rm{BKG}(v_{2}\{\psi\}).	
	\end{split}
	\label{EQ_RP7}
\end{equation}
By using the ZDC measured 1st order event plane $\psi_{1}$ as a estimation of the $\psiRP$, 
and 2nd order event plane $\psi_{2}$ reconstructed from final-state particles as a proxy of the $\psiEP$,
we can measure the $\dg\{\psiRP\}$ and $\dg\{\psiPP\}$. 
Assuming the $\rm{CME}(B_{sq}\{\psi\})$ are expected to be proportional to $B_{sq}$ and $\rm{BKG}(v_{2}\{\psi\})$ proportional to $v_{2}$, we have: 

\begin{equation}
	\begin{split}
		R_{EP}(\dg) = 2\frac{r(1-a^{EP}_{B_{sq}}) -(1-a^{EP}_{v_{2}})}{r(1+a^{EP}_{B_{sq}}) +(1+a^{EP}_{v_{2}})} \approx \frac{1-r}{1+r}R^{EP}(v_{2}).	
	\end{split}
	\label{EQ_RP8}
\end{equation}
Here $r \equiv \rm{CME}(B_{sq}\{\psiRP\})/\rm{BKG}(v_{2}\{\psiEP\})$ can be considered as the relative CME signal to background contribution,
\begin{equation}
	\begin{split}
		r = \frac{1+a^{EP}_{v_{2}}}{1+a^{EP}_{B_{sq}}} \frac{R^{EP}(\dg)-R^{EP}(v_{2})}{R^{EP}(B_{sq})-R^{EP}(\dg)} \approx \frac{R^{EP}(v_{2})-R^{EP}(\dg)}{R^{EP}(v_{2})+R^{EP}(\dg)}. 
	\end{split}
	\label{EQ_RP9}
\end{equation}
With respect to $\psiRP$ and $\psiEP$,  the CME signal fractions are, respectively,
\begin{equation}
	\begin{split}
		&f_{RP}(\rm{CME}) = \rm{CME}(B_{sq}\{\psiRP\})/\dg\{\psiRP\}=r/(r+\it{a}^{EP}_{v_{2}}),  \\ 
		&f_{EP}(\rm{CME}) = \rm{CME}(B_{sq}\{\psiEP\})/\dg\{\psiEP\}=r/(r+1/\it{a}^{EP}_{B_{sq}}).
	\end{split}
	\label{EQ_RP10}
\end{equation}

Experimentally, $R^{EP}(v_{2})$ can be estimated by $v_{2}$ measurements with respect to ZDC $\psi_{1}$ and second order event plane $\psi_{2}$ (such as the forward time projection chamber, FTPC). 
$R^{EP}(B_{sq})$ cannot but may be approximated by $−R^{EP}(v_{2})$, as demonstrated by the Monte Carlo Glauber calculations and AMPT (Fig.~\ref{FG_RP2}). 
Fig.~\ref{FG_RP3}(left) shows the STAR measured $\dg$ with respect to $\psi_{1}$ by ZDC and $\psi_{2}$ TPC\cite{Adamczyk:2013hsi,Abelev:2009ac,Abelev:2009ad}. 
Their relative difference can be used as a experimental estimation of the $R^{EP}(\dg) = 2(\dg\{\psi_{1}\} - \dg\{\psi_{2}\})/(\dg\{\psi_{1}\} - \dg\{\psi_{2}\} )$.
By assuming $a^{EP}_{v_{2}} = a^{EP}_{B_{sq}}$ and $R^{EP}(B_{sq}) = −R^{EP}(v_{2})$, 
the extracted fraction $f^{EP}_{CME}$ by Eq.~\ref{EQ_RP9},\ref{EQ_RP10} as a function of ``ture'' $v_{2}$ is shown in Fig.~\ref{FG_RP3}(right) 
by the thick curve as a function of the ``true'' $v_{2}$. 
The gray area is the uncertainty. The vertical lines indicate the various measured $v_{2}$ values. 
At present the data precision does not allow a meaningful constraint on $f_{CME}$; 
the limitation comes from the $\dg\{\psi_{1}\}$ measurement which has an order of magnitude
larger statistical error than that of $\dg\{\psi_{2}\}$. 
With tenfold increase in statistics, the constraint would be the dashed curves. 
This is clearly where the future experimental emphasis should be placed: 
larger Au+Au data samples are being analyzed and more Au+Au statistics are to be accumulated; 
ZDC upgrade is ongoing in the CMS experiment at the LHC; 
fixed target experiments at the SPS may be another viable venue where all spectator nucleons 
are measured in the ZDC allowing possibly a better determination of $\psi_{1}$\cite{Xu:2017qfs}.

%%%%%%%%%%%%%%%%%%%%%%%%%%%%%%%%%%%%%%%%%%%%%%%%%%%%%%%%%%%%%%%%%%%%%%%%%%%%%%%%%%%%%%%%%%%%%%
\section{Invariant mass method}
It has been known since the very beginning that the $\dg$ could be contaminated by background from resonance decays coupled with the elliptic flow ($v_{2}$)\cite{Voloshin:2004vk}.
Only recently, a toy-model simulation estimate was carried out which indicates 
that the resonance decay background can indeed largely account for the experimental measured $\dg$\cite{Wang:2016iov}, contradictory to early claims \cite{Voloshin:2004vk}. 
The pair invariant mass would be the first thing to examine in terms of resonance background, however, the 
invariant mass ($m_{inv}$) dependence of the $\dg$ has not been studied until recently\cite{Zhao:2017nfq}.
The invariant mass method of the $\dg$ measurements provides the ability to identify and remove resonance decay background, 
enhancing the sensitivity of the measured CME signal.

CME-driven charge separation refers to the opposite-sign charge moving in opposite directions along the magnetic field ($\Bvec$).
Because of resonance elliptic anisotropy ($\vres$),
more OS pairs align in the $\psiRP$ than $\Bvec$ direction, and it is an anti-charge separation along $\psiRP$.
This would mimic the same effect as the CME on the $\gdel$ variable\cite{Voloshin:2004vk,Abelev:2009ac,Abelev:2009ad}. In term of the $\dg$ variable, these backgrounds can be expressed by:

\begin{eqnarray}
	\gdel&\propto&\mean{\cos(\alpha+\beta-2\phires)\cos2(\phires-\psiRP)}\nonumber\\
			   &\approx&\mean{\cos(\alpha+\beta-2\phires)}\vres\,.
	\label{EQ_IM1}
\end{eqnarray}
where $\mean{\cos(\alpha+\beta-2\phires)}$ is the angular correlation from the resonance decay, $\vres$ is the $v_{2}$ of the resonance.
The factorization of $\mean{\cos(\alpha+\beta-2\phires)}$ with $\vres$ is only approximate, because both $\mean{\cos(\alpha+\beta-2\phires)}$ and $\vres$ depend on $\pt$ of the resonance.

\begin{figure}
	\centering 
	\includegraphics[width=13cm]{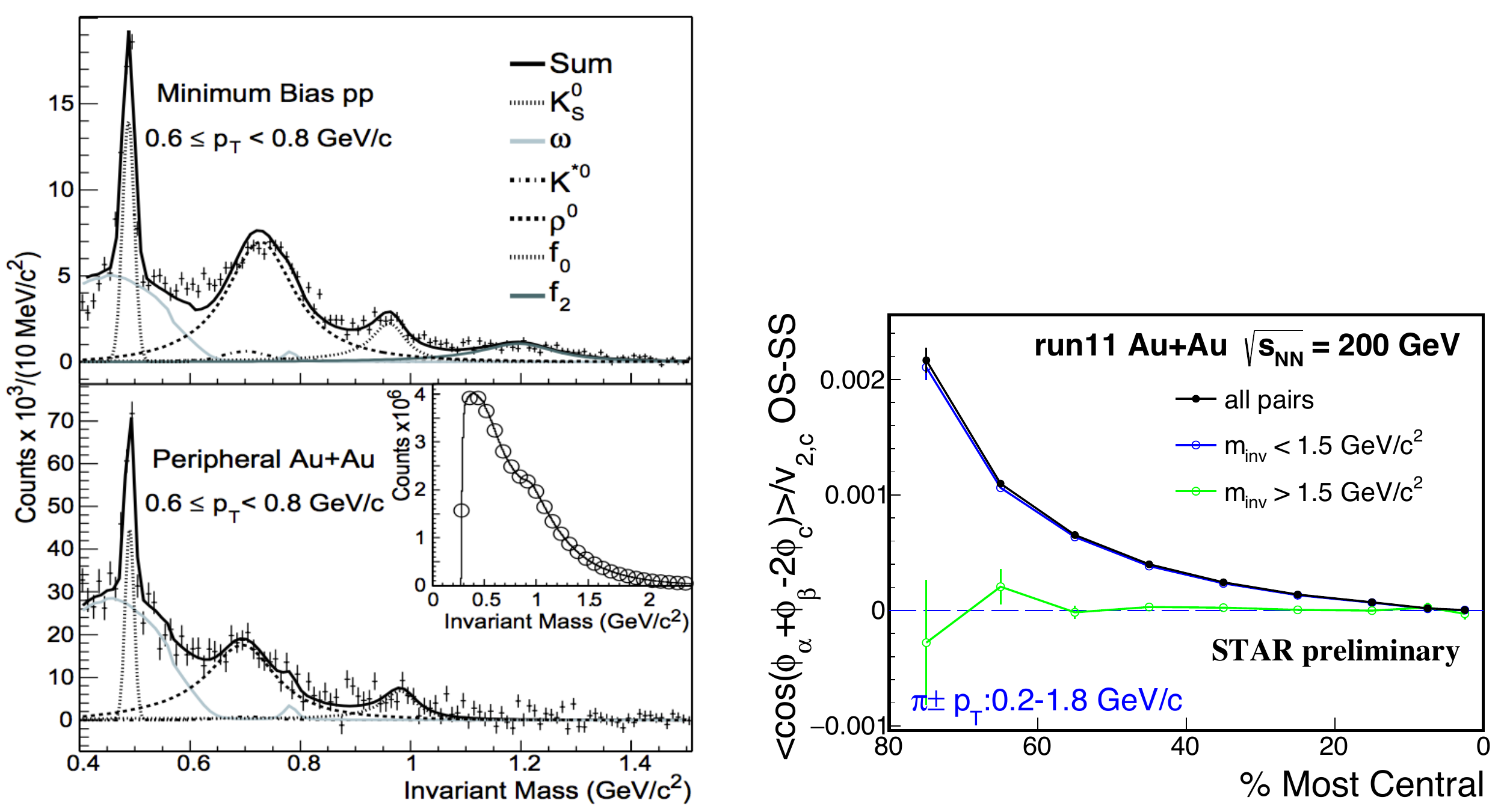}
	\caption{
       (Left panel) The raw $\pi$-$\pi$ invariant mass ($m_{inv}$) distributions after subtraction of the like-sign reference distribution for minimum bias p+p (top) 
       and peripheral Au+Au (bottom) collisions. 
       The insert plot corresponds to the raw $\pi^{+}$-$\pi^{-}$ $m_{inv}$ (solid line) and the like-sign reference 
       distributions (open circles) for peripheral Au+Au collisions\cite{Adams:2003cc}.
	   (Right panel) The inclusive $\gdel$ over all mass (black) and at $m_{inv} > 1.5$ \GeVcsq (green) as a function of centrality in Au+Au collisions at 200 GeV\cite{Zhao:2017wck}. 
	}
	\label{FG_IM1}
\end{figure}

Many resonances have broad mass distributions\cite{Agashe:2014kda}. 
Experimentally, they are hard to identify individually in relativistic heavy-ion collisions.
Statistical identification of resonances does not help eliminate their contribution to the $\dg$ variable.
However, most of the $\pi$-$\pi$ resonances contributions are dominated at low invariant mass region (Fig.~\ref{FG_IM1}, left)\cite{Adams:2003cc},
It is possible to exclude them entirely by applying a lower cut on the invariant mass, for example $m_{inv}>2.0$ \GeVcsq.
Results from AMPT model show that with such a $m_{inv}$ cut, although significantly reducing the statistics, 
can eliminate essentially all resonance decay backgrounds\cite{Zhao:2017nfq}.
The preliminary experimental data from STAR show similar results as AMPT. 
Fig.~\ref{FG_IM1}(right) shows the results with and without such an invariant mass cut.
By applying the mass cut, the $\dg$ is consistent with zero with current uncertainty in Au+Au collisions at 200 GeV\cite{Zhao:2017wck}.
The results are summarized in Table~\ref{TB_IM1}. 

\begin{table}
	\tbl{The preliminary experimental data on inclusive $\dg$ over all mass and $\dg$ at $m_{inv}>1.5$ \GeVcsq for different centralities in Au+Au collisions at 200 GeV\cite{Zhao:2017wck}.} 
	{\begin{tabular}{lccc}
			\toprule
			Centrality  &  $\dg$ in all mass (A) &  $\dg$ at $m_{inv}>1.5$ \GeVcsq (B) & B/A \\ 
			\colrule
			50-80\% & $(7.45\pm0.21)\times10^{-4}$   & $(1.3\pm5.7)\times10^{-5}$ & $(1.8\pm7.6)\%$ \\
			20-50\% & $(1.82\pm0.03)\times10^{-4}$   & $(7.7\pm9.0)\times10^{-6}$ & $(4.3\pm4.9)\%$ \\
			0-20\% & $(3.70\pm0.67)\times10^{-5}$   & $(-0.1\pm1.8)\times10^{-5}$ & $(-3.8\pm49)\%$  \\
			\botrule
		\end{tabular}
	\label{TB_IM1}}
\end{table}

While CME is generally expected to be a low $\pT$ phenomenon\cite{Kharzeev:2007jp,Abelev:2009ad}; its contribution to high mass may be small. 
In order to extract CME at low mass, resonance contributions need to be subtracted.
The invariant mass $\dg$ measurement provides such a tool that could possibly isolate the CME from the resonance background, 
by taking advantage of their different dependences on $m_{inv}$. 

For example, the $\rho$ decay background contribution to the $\dg$ is:
\begin{eqnarray}
	(\dg)_{\rho}& = \frac{N_{\rho}}{(N_{\pi^{+}+\pi^{-}})} \gamma_{\rho} = r_{\rho} \gamma_{\rho}  
	\label{EQ_IM2}
\end{eqnarray}
where $r_{\rho}$ is the relative abundance of $\rho$ decay pairs over all OS pairs, 
and $\gamma_{\rho} \equiv \mean{\cos(\alpha+\beta-2\phires)\vres}$ quantifies the $\rho$ decay angular correlations coupled with its $v_{2}$.
Consider the event to be composed of primordial pions containing CME signals (CME) and common (charge-independent) background, 
such as momentum conservation ($\gamma_{m.c.}$)\cite{Bzdak:2010fd,Pratt:2010zn}, 
and the resonance ($\rho$ for instance) decay pions containing correlations from the decay\cite{Voloshin:2004vk,Schlichting:2010qia,Wang:2016iov}. 
The $\minv$ dependency of the $\dg$ can be expressed as: 
\begin{equation}
	\begin{split}
		\dg(\minv) & = \frac{N_{SS}(\gamma_{CME}+\gamma_{m.c.}+N_{\rho}\gamma_{\rho})}{N_{SS}+N_{\rho}} - (-\gamma_{CME} + \gamma_{m.c.}) \\	
        & = r(\minv)(\gamma_{\rho} - \gamma_{m.c.}) + (1-r(\minv)/2)\dg_{CME}  \\ 
		& \approx r(\minv)R(\minv) + \dg_{CME}(\minv). 
	\end{split}
	\label{EQ_IM2}
\end{equation}
The first term is resonance contributions, where the response function $R(\minv) = \gamma_{\rho} - \gamma_{m.c.} = \mean{f(\minv)v_{2}(\minv)} - \gamma_{m.c.}$
is likely a smooth function of $\minv$, while $r(\minv)$ contains resonance spectral profile. 
Consequently, the first term is not smooth but a peaked function of $\minv$.
The second term in Eq.~\ref{EQ_IM2} is the CME signal which
should be a smooth function of $\minv$ (here the negligible $r/2$ term was dropped). 
However, the exact functional form of CME($\minv$) is presently unknown and
needs theoretical input. The different dependences of the two terms can be exploited to identify CME signals at low $\minv$.
The possibility of the this method was studied by a toy-MC simulation along with the AMPT models\cite{Zhao:2017nfq}.

\begin{figure}[htbp!]
	\centering 
	\includegraphics[width=7.0cm]{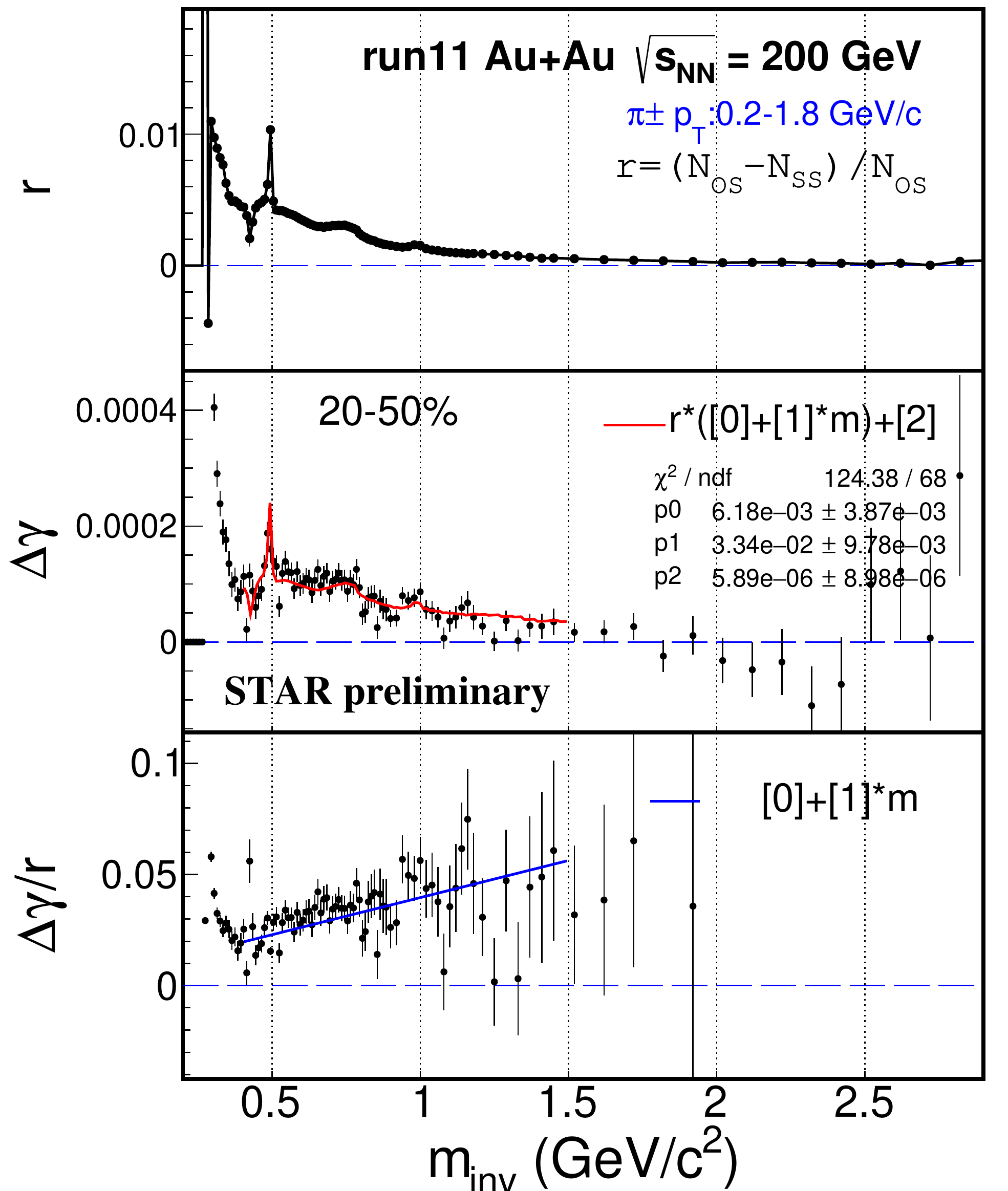} 
	\caption{(Color online)
		Pair invariant mass ($m_{inv}$) dependence of the relative excess of OS over SS charged $\pi$ pair multiplicity, $r=(N_{OS}-N_{SS})/N_{OS}$ (top panel), 
		event plane dependent azimuthal correlator difference, $\gdel=\gOS-\gSS$ (middle panel), and the ratio of $\gdel/r$ (bottom panel) in 20-50\% 
		Au+Au collisions at 200 GeV. Errors shown are statistical.
		The red curve in the middle panel shows the two-component model fit assuming a constant CME contribution independent of $m_{inv}$;
		The blue curve in the bottom panel shows the corresponding resonance response function\cite{Zhao:2017wck}.
	}   
	\label{FG_IM2}
\end{figure}

Figure~\ref{FG_IM2} shows the preliminary results in mid-central Au+Au collisions from STAR experiments\cite{Zhao:2017wck}.
Fig.~\ref{FG_IM2}(top) shows the relative OS and SS pair difference ($r=(N_{OS.}-N_{SS.})/N_{OS.}$) as a function of invariant mass.
Fig.~\ref{FG_IM2}(middle) shows the $\dg$ correlator as function of $\pi$-$\pi$ invariant mass. 
The data shows resonance structure in $\dg$ as function of mass; 
a clear resonance peak from $K_{s}^{0}$ decay are observed, and possible $\rho$ and $f^{0}$ peaks are also visible.
The $\dg$ correlator traces the distribution of those resonances. $\dg$ decreases as $r$ decreases with increasing mass,
In a two components model of resonances background plus CME signal. The $\dg(m)=r\times(a+b\times m)$+f(CME), where f(CME) represents the CME contribution.
The background contribution will follow the distribution of $r$, while the f(CME) is most likely a smooth distribution in $m_{inv}$.
Fig.~\ref{FG_IM2}(bottom) shows the ratio of the $\dg$/$r$ as function of mass.
No evidence of inverse shape of the resonance mass distribution is in the ratio of $\dg$/$r$, suggesting insignificant CME signal contributions.

In order to isolate the possible CME from the resonances contributions, 
the two components model is used to fit the $\dg$ as function of invariant mass (Fig.~\ref{FG_IM2} (middle)). 
Currently, there is no available theoretical calculation on the mass dependence of the CME contribution, 
therefore two functional forms are considered: (\rm{i}) a constant CME distribution independent of mass, and (\rm{ii}) a exponential CME distribution as function of mass.
The extracted $\dg$ from CME contribution is $(5.9\pm9.0)\times10^{-6}$ from the constant CME fit, 
and $(3.0\pm2.0)\times10^{-5}$ from the exponential CME fit,
which correspond to $(3.2\pm4.9)$\% (constant CME) and $(16\pm11)$\% (exponential CME) of the inclusive $\dg$ ($(1.82\pm0.03)\times10^{-4}$) measurement. 
The results are also summarized in Table~\ref{TB_IM2}. 
Future theoretical calculations of the CME mass dependence would help to understand the results more precisely.

\begin{table}
	\tbl{
		The preliminary experimental data of the average $\dg$ signal (corresponding to the CME contribution in the two-component fit model) 
		extracted from the model fit at $m_{inv}<1.5$ \GeVcsq in mid-central (20-50\%) Au+Au collisions at 200 GeV\cite{Zhao:2017wck}, 
		with two assumptions for the CME $m_{inv}$ dependence: a constant independent of $m_{inv}$ and an exponential in $m_{inv}$.
	}
	{\begin{tabular}{lcc}
			\toprule
			$\dg$ (inclusive)          &   \multicolumn{2}{c}{$(1.82\pm0.03)\times10^{-4}$}  \\ 
			\colrule
			& constant CME             & exponential CME in $m_{inv}$  \\
			average signal $\dg$ (fit) & $(5.9\pm9.0)\times10^{-6}$   & $(3.0\pm2.0)\times10^{-5}$ \\
			fit/inclusive              & $(3.2\pm4.9)\%$              & $(16\pm11)\%$ \\
			\botrule
		\end{tabular}
	\label{TB_IM2}}
\end{table}

Invariant mass method provides for the first time a useful tool to identify the background sources for the 
CME $\dg$ measurements, and provides a possible way to isolate the CME signal from the backgrounds. 
There are still debates weather the CME should be a low $\pT$/$\minv$ phenomenon, and their $\minv$ dependence is also not clear currently. 
Recent study\cite{Shi:2017cpu} indicates that the CME signal is rather independent of $\pT$ at $\pT>0.2$ \GeVc, 
suggesting that the signal may persist to high $\minv$.
Nevertheless, a lower $\minv$ cut will eliminate resonance contributions to $\dg$, 
and a measured positive $\dg$($\minv$) signal would point to the possible existence of the CME at
high $\minv$. A null measurement at high $\minv$, however, does not necessarily mean no CME also at low $\minv$. 
Further theoretical calculation on the CME $\minv$ dependence could help to extract the CME signal more precisely. 
On the other side, using ESE method to select events with different $v_{2}$ might be able to help to extract the background $\minv$ distributions 
by comparing their $\minv$ dependences of the $\dg$ distributions. 
In the upcoming isobar run at RHIC, it is also worthwhile to compare the $\dg$($\minv$) dependences between the two systems, 
which could help to understand where the possible CME $\dg$ signal comes from, 
for example the resonance abundance difference due to isospin difference between Zr and Ru or other effects\cite{Wang:2016iov}.
Further more it could also help to locate $\minv$ position of the possible CME $\dg$ signal and possibly provide the only way 
to study the $\minv$ property of the sphaleron or instanton mechanism for transitions between QCD vacuum states.

%%%%%%%%%%%%%%%%%%%%%%%%%%%%%%%%%%%%%%%%%%%%%%%%%%%%%%%%%%%%%%%%%%%%%%%%%%%%%%%%%%%%%%%%%%%%%%
\section{$R(\Delta S)$ correlator}
Recently a new observable, $\Rsm{m}$ (m=2, 3, refer to $\psi_{2}, \psi_{3}$), 
has been proposed to measure the CME-driven charge separation in heavy-ion collisions\cite{Magdy:2017yje,Ajitanand:2010rc}. 
\begin{equation}
	\begin{split}
		&\Delta S = \langle S_{p}^{h+} \rangle - \langle S_{n}^{h-} \rangle, \\ 
		&\langle S_{p}^{h+} \rangle = \frac{\displaystyle\sum_{1}^{N_{p}}\sin( \frac{m}{2} \Delta\phi_{m})}{N_{p}}, \langle S_{n}^{h-} \rangle = \frac{\displaystyle\sum_{1}^{N_{n}}\rm{sin}( \frac{m}{2} \Delta\phi_{m})}{N_{n}}, \\ 
		&\Delta\phi_{m} = \phi - \psi_{m},  m=2,3,
	\end{split}
\end{equation}
where $\phi$ is the azimuthal angle of the positively (p) or negatively (n) charged hadrons.
$\Delta S$ quantifies the charge separation along a certain direction. 
The correlation functions $\Cpm{m}$ were constructed from the ratio of the $N_{real}(\Delta S)$ distribution to the charge-shuffled $N_{shuffled}(\Delta S)$ distribution. 
\begin{equation}
	\begin{split}
		\Cpm{m} = \frac{N_{real}(\Delta S)}{N_{\rm{shuffled}}(\Delta S)}.
	\end{split}
\end{equation}
The $N_{\rm{shuffled}}(\Delta S)$ distribution was obtained by randomly shuffling the charges of the positively and negatively charged particles in each event. 
By replacing the $\psi_{m}$ with $\psi_{m} + \pi/m$, the same procedures were carried out to obtain the $\Cpmp{m}$.  
The $\pi/m$ rotation of the event planes, guarantees that a possible CME-driven charge separation does not contribute to $\Cpmp{m}$. 
In the end, the $\Rsm{m}$ correlator was obtained by taken the ratio between $\Cpm{m}$ and $\Cpmp{m}$:
\begin{equation}
	\begin{split}
		\Rsm{m} = \Cpm{m} / \Cpmp{m}.
	\end{split}
\end{equation}
The $\Cpm{m}$ measures the combined effects of CME-driven charge separation and the background, and the $\Cpmp{m}$ provides the reference for the background. 
The ratio between the $\Cpm{m}$ and $\Cpmp{m}$ are designed to detect the CME-driven charged separation. 

The CME-driven charge separation is along the magnetic field direction, which is perpendicular to the $\psiRP$. 
By using the $\psi_{2}$ as a proxy of the $\psiRP$, the $\Rsm{2}$ are designed to provide the sensitivity to detect the CME-driven charged separation. 
Since there is little, if any, correlation between $\psiRP$ and $\psi_{3}$, the $\Rsm{3}$ measurements
are insensitive to CME-driven charge separation, but still sensitive to background\cite{Magdy:2017yje}.

\begin{figure}[htbp!]
	\centering 
	\includegraphics[width=12.5cm]{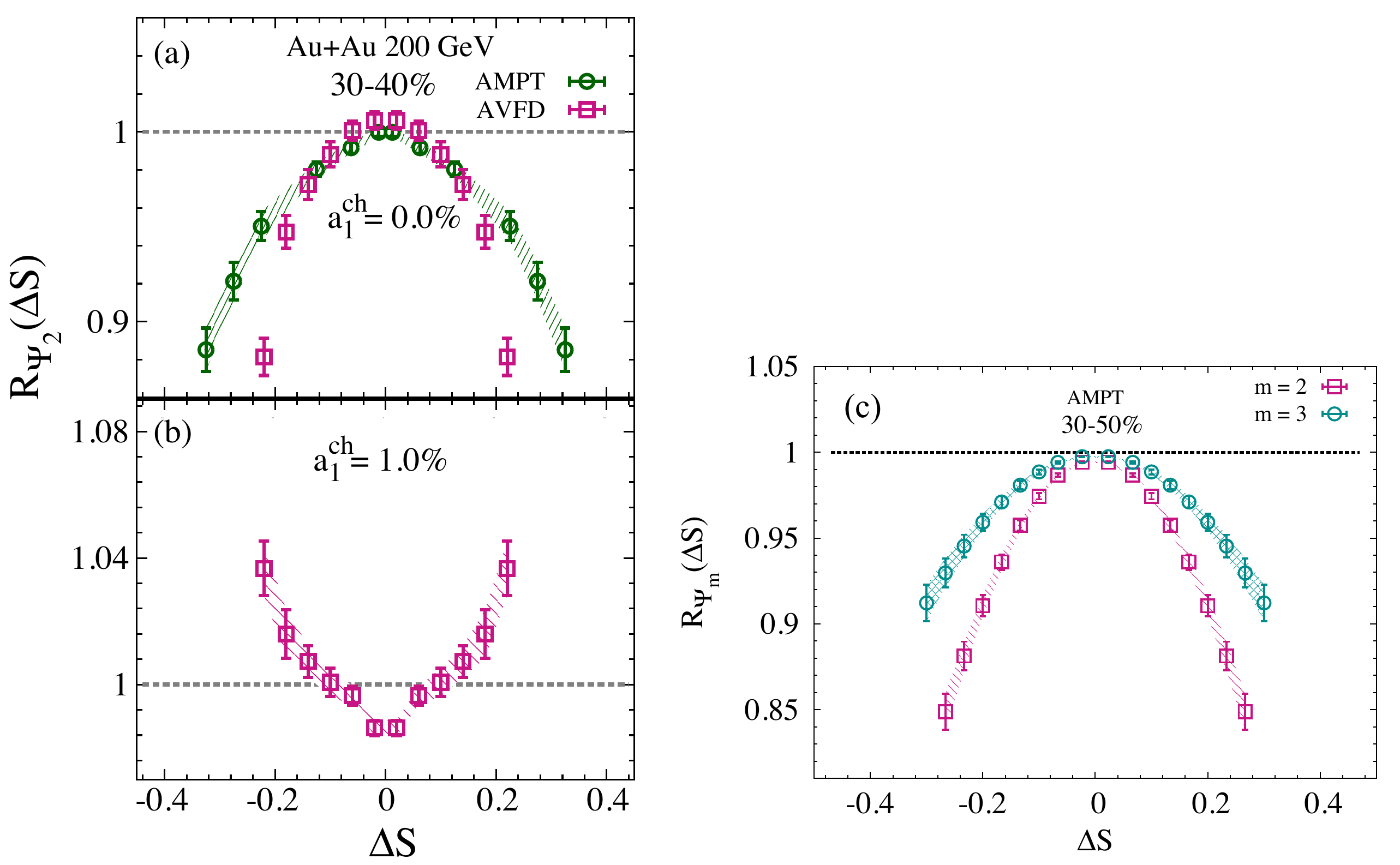} 
	\caption{(Color online)
		Comparison of the $\Rsm{2}$ correlators for (a) background-driven charge separation ($a_{1}$ = 0) 
		in 30-40\% Au+Au collisions (\sNN = 200 GeV) obtained with the
		AMPT and AVFD models, and (b) the combined effects of background- and CME-driven 
		($a_{1} = 1.0\%)$ charge separation in Au+Au collisions obtained with the AVFD model at
		the same centrality and beam energy.
		(c) Comparison of the $\Rsm{3}$ correlators for background-driven charge separation ($a_{1}$ = 0) 
		in 30-50\% Au+Au collisions (\sNN = 200 GeV) obtained with the AMPT model\cite{Magdy:2017yje}.
	}
	\label{FG_RS1}
\end{figure}

Figure~\ref{FG_RS1} shows the initial studies with A Multi-Phase Transport (AMPT) and Anomalous Viscous Fluid Dynamics (AVFD) models\cite{Magdy:2017yje}.
The AMPT\cite{Zhang:1999bd,Lin:2004en} has been quite successful in describing the experimentally measured data (particle yields, flow) in heavy ion collisions.  
Therefore it provides a good reference for the background response of the $\Rsm{m}$ correlator, especially the resonance decay and the flow related background. 
In additional to the background, the AVFD model\cite{Shi:2017cpu} could include the evolution of chiral fermion currents in the hot dense medium during the bulk hydrodynamic evolution. 
which can be used to study the $\Rsm{m}$ response to the CME-driven charge separation.
Both the AMPT and AVFD shows the convex shapes of $\Rsm{2}, \Rsm{3}$ for typical resonance backgrounds (Fig.~\ref{FG_RS1} panel (a,c)). 
With implementing anomalous transport from the CME, the AVFD model simulation shows a concave $\Rsm{2}$ distribution (Fig.~\ref{FG_RS1} panel (b)), 
which is consistent with the expectation of the $\Rsm{2}$ correlator response to the CME-driven charge separation. 
Preliminary experimental data from STAR, reveal concave $\Rsm{2}$ distributions in 200~GeV Au+Au collisions\cite{Roy:2017rs}.

A 3+1-dimensional hydrodynamic study\cite{Bozek:2017plp}, however, indicates concave $\Rsm{2}$ shapes for backgrounds as well, 
it also shows a concave shapes of $\Rsm{3}$ 
distribution for the background, which is different from the expectation of convex shape of background.

To better understand those results from different models, hence gain more information from the experimental data, 
a more detailed and systematic study of $\Rsm{m}$ correlator responses to the background seems important. 
For example, the $\Rsm{3}$ response in AVFD model with and without CME-driven charge separation.
And the resonance $v_{n}$, and $p_{T}$ dependences of the $\Rsm{m}$ behavior\cite{Feng:2018chm}. 
The resonance $v_{2}$ introduces different numbers of decay $\pi^+\pi^-$ pairs in the in-plane and out-of-plane directions.
The resonance $p_{T}$ affects the opening angle of the decay $\pi^+\pi^-$ pair.
Low $p_{T}$ resonances decay into large opening-angle pairs, and result in more ``back-to-back'' pairs out-of-plane because of the more in-plane resonances,
mimicking a CME charge separation signal perpendicular to the reaction plane, or a concave $\Rsm{2}$.
High $p_{T}$ resonances, on the other hand, decay into small opening-angle pairs, and result in a background behavior of convex $\Rsm{2}$.

Other than the $\dg$ correlator, it is worth developing new methods and/or observables to search for the CME, such as the $\Rsm{m}$ correlator. 
Currently more detailed investigations are needed to understand how the $\Rsm{m}$ correlator is compared with other correlators, 
what the advantage and disadvantage of the $\Rsm{m}$ correlator is, and possibly what the connections are between these correlators.
More detailed studies could help gain a better understanding of the experimental results and more clear interpretation in term of CME, 
and future study of the RHIC isobaric data \cite{Magdy:2018lwk,Sun:2018idn}.

%%%%%%%%%%%%%%%%%%%%%%%%%%%%%%%%%%%%%%%%%%%%%%%%%%%%%%%%%%%%%%%%%%%%%%%%%%%%%%%%%%%%%%%%%%%%%%
\section{summary}
The non-trivial topological structures of the QCD have wide ranging implications.
Relativistic heavy-ion collisions provide an ideal environment to study the novel phenomena induced by those topological structures, 
such as the chiral magnetic effect (CME). 
Since the first $\gamma$ measurements in 2009,
experimental results have been abundant in relativistic heavy-ion as well as small system collisions.
In this review, several selected recent progresses on the experimental search for the CME in relativistic heavy-ion collisions are summarized.
Major conclusions are as follows:

\begin{itemlist}
\item \textbf{Event shape selection}: Using the event shape selection, by varying the event-by-event $v_{2}$, exploiting statistical (event-by-event $v_{2}, q_{2}$ methods) and dynamical fluctuations (ESE method),
experimental results suggest that the $\dg$ correlator is strongly dependent on the $v_{2}$.
The $v_{2}$ independent contribution are estimated by different methods from STAR, ALICE and CMS collaboration;
results indicate that a large contribution of the $\dg$ correlator is from the $v_{2}$ related background.
\item \textbf{Isobaric collisions and Uranium+Uranium collisions}: By taking advantage of the nuclear property 
(such as proton number, shape), 
isobaric collisions of $\ZrZr$, $\RuRu$ collisions and Uranium+Uranium collisions have been proposed. 
So far there is no clear conclusion in term of the disentangle of the CME and $v_{2}$ related background from the preliminary experimental Uranium+Uranium results yet. 
Theoretical calculations suggest that the upcoming isobaric collisions at RHIC in 2018 will provide a powerful tool to disentangle the CME signal from the $v_{2}$ related backgrounds.
While there could be non-negligible deviations of the $\Ru$\ and $\Zr$ nuclear densities from Woods-Saxon which could introduce extra uncertainty. 
\item \textbf{Small system collisions}: 
The recent $\dg$ measurements in small system p+Pb collisions from CMS have triggered a wave of 
discussions about the interpretation of the CME in heavy-ion collisions.
Preliminary results from STAR also show comparable $\dg$ in small system \pdAu\ collisions with that in \AuAu\ collisions. 
These results indicate significant background contributions in the $\dg$ measurements in heavy-ion collisions. 
On other hand, theoretical calculation shows a possibility that CME may contribute to the $\dg$ in p+Pb collisions with respect to $\psi_{2}$.
The $\dg$ measurements in small system \pdA\ collisions with respect to $\psi_{1}$ using the spectator neutrons are worth to follow in the future. 
\item \textbf{Measurement with respect to the reaction plane}: New idea of differential measurements with respect to the reaction plane ($\psiRP$) and participant plane ($\psiPP$) are proposed, 
where the $\psiRP$ could possibly be assessed by spectator neutrons measured by the zero-degree calorimeters (ZDC). 
The $v_{2}$ is stronger along $\psiPP$ and weaker along $\psiRP$; in contrast, the magnetic field, being from spectator protons, is weaker along
$\psiPP$ and stronger along $\psiRP$. The $\dg$ measured with respect to $\psiRP$ and $\psiPP$ contain different amounts of CME and background, 
and can thus determine these two contributions. 
\item \textbf{Invariant mass method}: 
New method exploiting the invariant mass dependence of the $\dg$ measurements provides a useful tool to identify the background sources, 
and provides a possible way to isolate the CME signal from the backgrounds.
Preliminary results from STAR show that by applying a mass cut to remove the resonance background, the $\dg$ is consistent with zero with current uncertainty in Au+Au collisions.
In the low mass region, resonance peaks are observed in $\dg$ as a function of $\minv$. 
By assuming smooth CME $\minv$ distribution, it's possible to extract the CME signal.
While there are debates wheather the CME should be a low $\pT$/$\minv$ phenomenon, their $\minv$ dependence is also not clear currently.
In the upcoming isobar run at RHIC, the comparison of the $\dg$($\minv$) dependences between the two systems would help to further our understanding.
and will provide a possible way to study the $\minv$ property of the sphaleron or instanton mechanism for transitions between QCD vacuum states.
\item \textbf{$R(\Delta S)$ correlator}:
New $R(\Delta S)$ correlator has been proposed to measure the CME-driven charge separation. 
Preliminary experimental results indicate a CME dominated scenario. 
To gain better understanding of the experimental results and more clear implications in term of its CME interpretation, 
more detailed investigations are needed, such as, the resonance $v_{n}$, and $p_{T}$ dependences of the $R(\Delta S)$ behavior.
\end{itemlist}

While the physics behind CME is of paramount importance, the present experimental evidences for the existence of the CME are rather ambiguous. 
Most of the results indicate that there are significant background contributions in the $\dg$ measurements, 
the CME signal might be small fraction, while there is no doubt that the unremitting pursuit is encouraging and will be rewarded. 
Toward the discovery of the CME, new ideas, new methods, new technologies are called for. 
The author is hopeful that this day will come soon.

%%%%%%%%%%%%%%%%%%%%%%%%%%%%%%%%%%%%%%%%%%%%%%%%%%%%%%%%%%%%%%%%%%%%%%%%%%%%%%%%%%%%%%%%%%%%%%
\section*{Acknowledgments}
I greatly thank Prof. Fuqiang Wang, Prof. Wei Xie and other members of the Purdue High Energy Nuclear Physics Group for discussions and comments.
This work was supported by the U.S. Department of Energy (Grant No. de-sc0012910).

%%%%%%%%%%%%%%%%%%%%%%%%%%%%%%%%%%%%%%%%%%%%%%%%%%%%%%%%%%%%%%%%%%%%%%%%%%%%%%%%%%%%%%%%%%%%%%
\bibliographystyle{ws-ijmpa}
\bibliography{ref}

\end{document}